\newcommand{\mc}{\multicolumn}
\documentstyle{mn}
\newcommand{\ltsimeq}{\raisebox{-0.6ex}{$\,\stackrel 
        {\raisebox{-.2ex}{$\textstyle <$}}{\sim}\,$}}

\begin{document}

\title[Optical spectroscopy of the 6CE sample]
{A sample of 6C radio sources with virtually
complete redshifts. II -- optical spectroscopy}

\author[Rawlings et al.]{Steve Rawlings$^{1\star}$, 
Steve Eales$^{2}$ and Mark Lacy$^{1,3,4}$ \\
$^1$Astrophysics, Department of Physics, Keble Road, Oxford OX1 3RH, UK \\
$^{2}$Department of Physics and Astronomy, University of Wales 
College of Cardiff, P.O. Box 913, Cardiff CF2 3YB, UK\\
$^{3}$Institute of Geophysics and Planetary Physics, L-413 Lawrence
Livermore National Laboratory, Livermore CA 94550, USA\\
$^{4}$Department of Physics, University of California, 1 Shields Avenue,
Davis CA 95616, USA\\
}
     
\maketitle

\begin{abstract}
\noindent
This is the second of two papers presenting
basic observational data on the 6CE sample of extragalactic
radio sources. It presents the results of optical spectroscopy 
which has yielded virtually complete redshift information for the
6CE sample: 56 of the 59 sample members have spectroscopic redshifts 
which, with the exception of seven cases, are secure. 
The redshift distribution $N(z)$ is fairly flat over the redshift 
range $0 \leq z < 2$, with a median redshift of $\approx 1.1$ 
and a high-redshift tail reaching to $z = 3.4$. The highest-redshift 
($z > 1.75$) members of the 6CE sample have similar optical spectra 
and a tight (less than one dex) spread in narrow-Ly$\alpha$ emission 
line luminosity.

\end{abstract}

\begin{keywords}
radio continuum:$\>$galaxies -- galaxies:$\>$active -- quasars:$\>$general 
\end{keywords}

\footnotetext{$^{\star}$Email: s.rawlings1@physics.ox.ac.uk}

\section{Introduction}
\label{sec:intro}

\begin{figure*}
\begin{center}
\setlength{\unitlength}{1mm}
\begin{picture}(150,120)
\put(-50,-40){\includegraphics{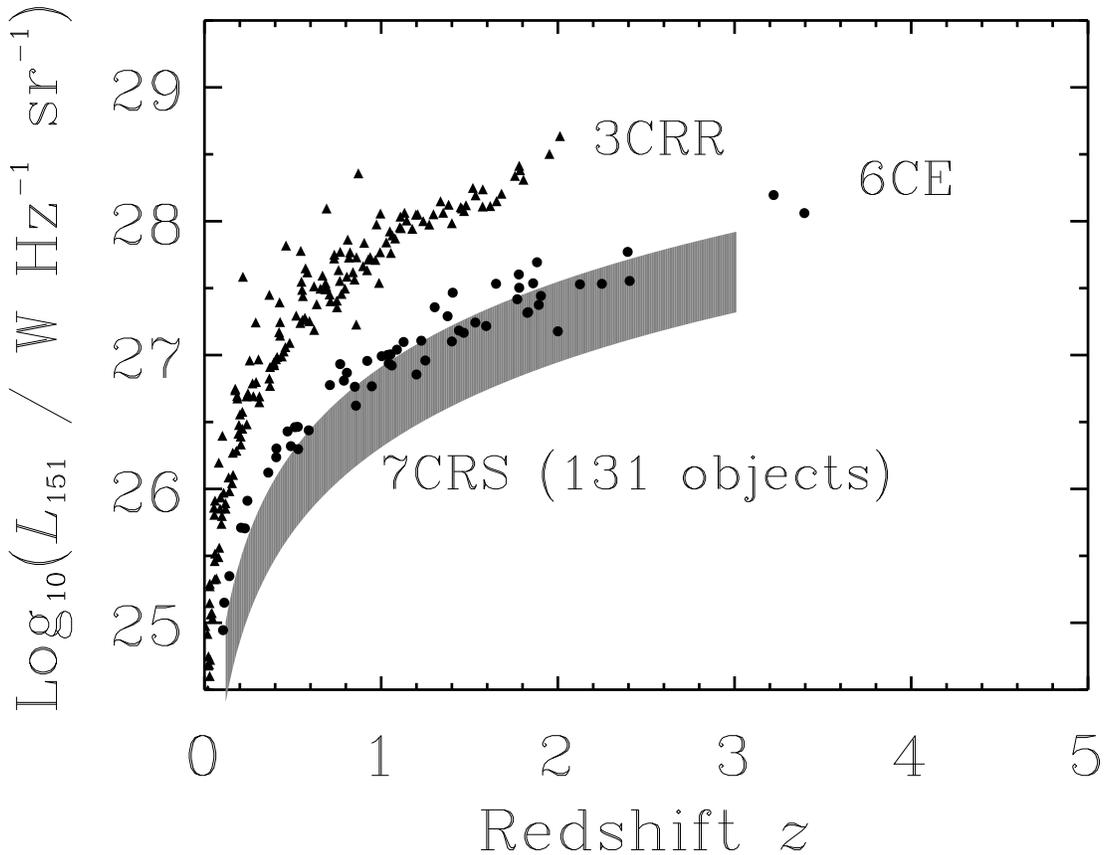}}
\end{picture}
\end{center}
{\caption[junk]{\label{fig:pz} The coverage of the 151 MHz luminosity 
$L_{151}$ versus redshift $z$ plane for the 3CRR sample (triangles)
and the 6CE sample (circles); the shaded region shows the rough location
of the 131 members of the 7C Redshift Survey (7CRS; Lacy et al.\ 1999;
Blundell et al.\ in prep.; Willott et al.\ in prep).
}}
\end{figure*}

Obtaining spectroscopic redshifts for complete samples
of radio sources is a key requirement for
a large number of investigations of the
radio source population and its cosmic evolution. Such investigations
are severely compromised 
if they are based on a sample selected at a single
limiting flux density, for example the 3CRR sample 
(Laing, Riley \& Longair 1983), because of the limited
coverage they provide of the 151 MHz luminosity $L_{151}$ versus
redshift $z$ plane (e.g.\ Fig.~\ref{fig:pz}). Studies based only on the
3CRR sample cannot distinguish between correlations with redshift $z$ and
correlations with $L_{151}$, and they cannot straightforwardly measure the 
relative numbers of radio sources at a given redshift as a function of
$L_{151}$, or conversely the relative numbers at a given $L_{151}$
as a function of $z$.

To ameliorate this problem we have been seeking
complete redshift information for low frequency (151 MHz) 
selected samples drawn from the 6C and 7C radio surveys
(e.g.\ Rawlings et al.\ 1998). These samples greatly increase 
the coverage of the $L_{151}$-$z$ plane as illustrated in 
Fig.~\ref{fig:pz}. In this paper we report on optical
spectroscopy of a 6C-based sample first defined by Eales (1985a) and now,
as described in this paper, slightly revised.
This is the second of two papers aimed at presenting
basic observational data on this revised `6CE' sample. The first paper
(Eales et al.\ 1997) presented near-IR images
of many members of the sample. 

A numbers of papers have made use of the 6CE dataset. Investigations of 
the inter-correlations between radio properties (e.g.\ luminosity, 
projected linear size, spectral index) and redshift can be found in 
Neeser et al.\ (1995), Best et al.\ (1999) and Blundell, Rawlings \& 
Willott (1999). Investigations of the near-IR properties of the 
combined 6CE/3CRR dataset can be found in Eales et al.\ (1997)
and Roche, Eales \& Rawlings (1998). An investigation of the
fraction of quasars in low frequency selected complete samples
can be found in Willott et al.\ (2000b). Measurements of the radio luminosity 
function using the combined 3CRR/6CE/7CRS dataset are 
presented in Willott et al.\ (2000c).

Throughout this paper we adopt the following values for
the cosmological parameters:
$H_{\circ}=50~ {\rm km~s^{-1}Mpc^{-1}}$, $\Omega_ {\rm M}=1$ and 
$\Omega_ {\Lambda}=0$.
B1950.0 coordinates are used and the convention for radio spectral index 
$\alpha$ is that $S_{\nu} \propto \nu^{-\alpha}$, where
$S_{\nu}$ is the flux density at frequency $\nu$.

\section{The 6CE sample}

\subsection{Sample membership}
\label{sec:sample}

Since the definition of the original 6C sample by Eales (1985a), 
the flux density scale of the 6C survey has been revised, 
and a number of programmes of follow-up at radio wavelengths 
(Naundorf et al.\ 1992; Law-Green et al.\ 1995a), and in the near-IR
and optical (e.g.\ Eales et al.\ 1997; this paper), have allowed us to 
make a proper assessment of source confusion. As a result, 
the 151 MHz flux density $S_{151}$ limits of the sample have been 
slightly altered, and the sample membership has been refined. The 
final selection criteria of the 6CE sample are as follows:

\begin{itemize}
\item $08^{\rm h}20^{\rm m} 30^{\rm s} <$ RA (B1950) $< 13^{\rm h}01^{\rm m}
30^{\rm s}$;

\item $34^\circ <$ Dec (B1950) $< 40^\circ$;

\item 2.00\,Jy $\leq S_{151} <$ 3.93\,Jy.

\end{itemize}

The sky area of the sample is $0.102 ~ \rm sr$. The 
region of sky was chosen initially by Eales (1985a) because of overlap with a
sample of (408-MHz selected) B2 radio sources
studied by Allington-Smith (1982) and it consisted of 67 sources.
Eight of these sources are excluded from the (revised) 
6CE sample: 6C0825+36, 6C0848+34, 6C0918+36, 6C0955+39
and 6C1106+38 because revised estimates of $S_{151}$
place the 6C catalogue entry below the flux-density limit
(Hales, Baldwin \& Warner 1988; Hales, Baldwin \& Warner 1993); and 
6C0951+37, 6C1028+35 and 6C1045+35C because the (integrated) $S_{151}$
6C catalogue entry results from
the confusion of two or more radio sources which, if considered
individually, would fall below the flux-density limit (see
Law-Green et al.\ 1995a). There are no new sources in the 6CE sample
which therefore comprises 59 radio sources. A summary of key information 
on the 6CE sample is provided in Table~\ref{tab:summary}; versions of this
table in both ASCII and HTML formats are available from the
worldwide-web at {\bf http://www-astro.physics.ox.ac.uk/}$\sim$
{\bf sr/6ce.html}.

One of the 59 6CE sources (6C1036+3616) is projected
so close to a bright star that any effective optical/near-IR
follow-up has so far proved impossible. Of the remaining 58 sources, five are `flat-spectrum'  
in the sense that their low-frequency radio spectra have indices lower
than 0.4 but it is not yet proven that any of these have been promoted into the
sample on the basis of Doppler-boosted components.

\scriptsize
\begin{table*}
\begin{center}
\begin{tabular}{rllrlllcrr}
\hline\hline
(1) & (2) & (3) & (4) & (5) & ~~~~(6) & (7) & (8) & (9) & (10)\\
\hline
& name & \mc{1}{c|}{$S_{151}$}
& \mc{1}{c|}{$\alpha_{151}$} & \mc{1}{c|}{Cl} & \mc{1}{c|}{$K$} & 
\mc{1}{c|}{$z$} & \mc{1}{c|}{Line} & \mc{1}{c|}{$\log_{10} L_{\rm line}$} &
\mc{1}{r|}{$z$ ref} \\
\hline

 & 6C0820+3642& 2.39 & 0.82 & HEG & $\phantom{>}18.41$ ~ (Eea  5)& 1.860      & Ly$\alpha$ &   36.55  & REL        \\    
A& 6C0822+3417& 3.06 & 0.55 & LEG & $\phantom{>}15.26$ ~ (LLA 12)& 0.406      & [OII]      &(  35.1  )& ALL        \\       
 & 6C0822+3434& 2.93 & 1.19 & LEG?& $\phantom{>}17.71$ ~ (Eea  5)& 0.768      & [OII]      &   34.78  & DSpc       \\       
A& 6C0823+3758& 3.35 & 0.76 & LEG & $\phantom{>}13.74$ ~ (LLA 12)& 0.207      & [OII]      &($<34.2 $)& ALL        \\       
A& 6C0824+3535& 2.42 & 0.41 & Q   &        -         & 2.249      &            &          & ASDL       \\
A& 6C0825+3452& 2.10 & 0.58 & HEG & $\phantom{>}19.19$ ~ (Eea  5)& 1.467      & [OII]      &   35.73  & EHRpc      \\       
A& 6C0847+3758& 3.07 & 0.91 & HEG & $\phantom{>}15.05$ ~ (LLA 12)& 0.407      & [OII]      &(  34.8  )& ALL        \\       
A& 6C0854+3956& 2.92 & 0.65 & HEG & $\phantom{>}15.39$ ~ (LLA 12)& 0.528      & [OII]      &(  35.3  )& ALL        \\       
A& 6C0857+3907& 2.71 & 0.73 & HEG & $\phantom{>}15.03$ ~ (LLA 12)& 0.229      &            &(  35.1  )& ALL        \\       
 & 6C0901+3551& 2.07 & 0.74 & HEG & $\phantom{>}18.29$ ~ (Eea  5)& 1.904      & Ly$\alpha$ &   36.67  & REW$\dag$  \\

\hline
 
A& 6C0902+3419& 2.14 & 0.80 & HEG & $\phantom{>}19.70$ ~ (Eb  4) & 3.395      & Ly$\alpha$ &   37.00  & L          \\
 & 6C0905+3955& 2.82 & 1.13 & HEG & $\phantom{>}18.48$ ~ (Eea  5)& 1.882      & Ly$\alpha$ &   36.31  & REL        \\
A& 6C0908+3736& 2.33 & 0.53 & LEG & $\phantom{>}12.70$ ~ (LLA 12 & 0.105      & [OII]      &($<34.2 $)& Wea        \\       
A& 6C0913+3907& 2.27 & 0.24 & Q   &        -         & 1.250      &            &          & La/ASDL  \\       
 & 6C0919+3806& 2.72 & 0.91 & LEG & $\phantom{>}18.20$ ~ (Eea  5)& 1.650?     &  CII]?     &   34.96  & REL$\ddag$ \\
A& 6C0922+3640& 3.27 & 0.58 & LEG & $\phantom{>}12.60$ ~ (LLA 12)& 0.112      & [OII]      &($<34.5 $)& S          \\
 & 6C0930+3855& 2.21 & 0.86 & LEG?& $\phantom{>}19.30$ ~ (ER   4)& 2.395      & Ly$\alpha$ &   36.10  & REL        \\   
 & 6C0943+3958& 2.31 & 0.82 & LEG?& $\phantom{>}18.09$ ~ (Eea  5 & 1.037      & [OII]      &   35.65  & REL        \\       
A& 6C0955+3844& 3.45 & 0.84 & Q   &        -         & 1.405      &            &          & La/ASDL  \\       
A& 6C1011+3632& 2.10 & 0.74 & HEG & $\phantom{>}18.05$ ~ (Eea  5)& 1.043      & [OII]      &   35.00  & REL        \\       

\hline

A& 6C1016+3637& 2.28 & 0.49 & HEG &$>20.30$~ (Eea  5)& 1.892      & Ly$\alpha$ &   36.01  & REL        \\
A& 6C1017+3712& 2.68 & 0.62 & HEG & $\phantom{>}18.88$ ~ (Eea  5)& 1.053      & [OII]      &   35.48  & REL        \\       
A& 6C1018+3729& 2.52 & 1.00 & HEG & $\phantom{>}16.49$ ~ (LLA 12)& 0.806      & [OII]      &   35.48  & ASDL       \\       
A& 6C1019+3924& 2.99 & 0.65 & LEG?& $\phantom{>}16.64$ ~ (LLA 12)& 0.921      & [OII]      &(  35.5)  & ALL        \\       
A& 6C1025+3900& 2.97 & 0.54 & LEG?& $\phantom{>}14.48$ ~ (LLA 12)& 0.361      & [OII]      &   35.46  & ASDL       \\       
 & 6C1031+3405& 2.33 & 0.47 & HEG & $\phantom{>}19.30$ ~ (ER   4)& 1.832      & Ly$\alpha$ &   36.20  & REL$\ddag$ \\       
A& 6C1036+3616& 2.81 &     -&    -&        -         &      -     &           -&         -& -            \\  
A& 6C1042+3912& 2.68 & 0.52 & HEG & $\phantom{>}19.70$ ~ (Eea  5)& 1.770      & Ly$\alpha$ &   36.26  & REL        \\
A& 6C1043+3714& 2.62 & 0.83 & LEG?& $\phantom{>}17.38$ ~ (LLA 12)& 0.789      & [OII]      &(  36.2  )& ALL        \\      
 & 6C1045+3403& 2.00 & 0.59 & HEG & $\phantom{>}18.91$ ~ (Eea  5)& 1.827      & Ly$\alpha$ &   36.77  & REL        \\

 \hline\hline
 \end{tabular}
 {\caption[Table 1]{\label{tab:summary} A summary of key information 
on the 6CE sample; the reason for the lack of data on 
6C1036+3616 is explained in Sec.~\ref{sec:sample}.
{\bf Column 1:} `A' indicates that the radio source is also in the B2-selected 
`1-Jy' sample of Allington-Smith (1982).
{\bf Column 2:} name of the 6CE radio source.
{\bf Column 3:} 151-MHz flux density from the final versions
of the 6C survey (Hales et al.\ 1998; Hales et al.\ 1993), 
and {\bf column 4:} radio spectral index evaluated at rest-frame 151 MHz using a polynomial fit 
to the multi-frequency data.
{\bf Column 5:} classification, Q=quasar and BLRG=broad-line radio galaxy, 
following the prescription of Willott et al.\ (1998); 
and HEG=high-excitation galaxy, LEG=low-excitation galaxy, 
following the prescription of Jackson \& Rawlings (1997)
when [OIII] is accessible, and using the detection of any
line from an at least doubly-ionized ion as a diagnostic 
in higher-redshift cases;
`?' means that the classification is uncertain -- 
object-by-object discussions 
are presented in Secs.~\ref{sec:obj_by_obj} and
~\ref{sec:obj_by_obj_lit}.
{\bf Column 6:} $K-$band magnitude of the radio source ID -- a reference to the
source of this magnitude and the diameter (in arcsec) of the photometric 
aperture,
is given in the brackets; `?' indicates that this $K-$band detection 
is significant at only the $2.6 \sigma$ level; `-' signifies that no
$K-$band photometry is available;
references are Ea=Eales et al.\ (1993a), Eb=Eales et al.\ (1993b),
Eea = Eales et al.\ (1997), ER = Eales \& Rawlings (1996),
LLA=Lilly, Longair \& Allington-Smith (1985), L89=Lilly (1989).
{\bf Column 7:} redshift, `?' means that this value is 
not yet an unequivocal redshift;
`-' indicates that spectroscopy has yet to reveal a redshift;
`*' indicates that the galaxy with this redshift is not 
yet certain to be the correct radio source ID.
{\bf Column 8:} Prominent emission line in the existing
spectra, a `?' meaning that the line identification is uncertain.
{\bf Column 9:} log$_{10}$ of the line luminosity in units of
W, a `-' means the data are inadequate to obtain a line luminosity; 
brackets denote that the luminosity is estimated from
an observed equivalent width (EW) and an $r-$ or $R-$band magnitude
(from either LLA or Eales 1985c); to estimate 
[OII] limits from the spectra of the low-redshift LEGs 
studied by Sargent (1973) or Warner et al.\ (1983),
we took $EW < 25 ~ \rm \AA$ and $EW < 10 ~ \rm \AA$ respectively, a detected
[OII] line in a Sargent spectrum is assigned
an $EW = 25 ~ \rm \AA$.
{\bf Column 10:}
reference to spectroscopy/spectrophotometry:
Apc, J. Allington-Smith {\it priv. comm.};
ALL, Allington-Smith et al.\ (1985); ASDL, Allington-Smith et al. (1988);
DSpc, M. Dickinson \& H. Spinrad {\it priv. comm.};
EH, Eracleous \& Halpern (1994);
EHRpc, S. Eales, C. Haniff \& S. Rawlings {\it priv. comm.};
ERWpc, S. Eales, S. Rawlings \& S. Warren {\it priv. comm.}; 
La, Lahulla et al. (1991);
L, Lilly (1988); 
LMRSpc, M. Lacy, S. Maddox, S. Rawlings \& S. Serjeant {\it priv. comm.};
REL, this paper; REW, Rawlings, Eales \& Warren (1990);
S, Sargent (1973); VT, Vermeulen \& Taylor (1995);
Wea, Warner et al. (1983); WW, Wills \& Wills (1979).
A $\dag$ indicates that the line flux has been measured from
a new spectrum presented in this paper. A 
$\ddag$ means that the previously reported redshift is
incorrect (see object-by-object notes in
Secs.~\ref{sec:obj_by_obj} and~\ref{sec:obj_by_obj_lit}).
 }}
 \end{center}
 \end{table*}

\normalsize

\addtocounter{table}{-1}

\clearpage

\scriptsize

\begin{table*}
\begin{center}
\begin{tabular}{lllrlllrlr}
\hline\hline
(1) & (2) & (3) & (4) & (5) & (6) & (7) & (8) & (9) & (10) \\
\hline
& name & \mc{1}{c|}{$S_{151}$}
& \mc{1}{c|}{$\alpha_{151}$} & \mc{1}{c|}{Cl} & \mc{1}{c|}{$K$} & 
\mc{1}{c|}{$z$} & \mc{1}{c|}{Line} & \mc{1}{c|}{$\log_{10} L_{\rm line}$} &
\mc{1}{r|}{$z$ ref.} \\

\hline

 & 6C1045+3553& 2.07 & 0.76 & LEG?& $\phantom{>}16.33$ ~ (Eea  5)& 0.851?     & [OII]?     &(  35.4  )& LMRSpc     \\
 & 6C1045+3513& 3.03 & 0.20 & Q   & $\phantom{>}16.77$ ~ (Eea  5)& 1.594      &           -&          & ERWpc$\ddag$ \\       
A& 6C1100+3505& 2.26 & 0.58 & HEG & $\phantom{>}18.33$ ~ (Eea  5)& 1.440      & [NeIV]     &   35.73  & Apc        \\       
A& 6C1108+3956& 2.10 & 0.74 & LEG & $\phantom{>}16.27$ ~ (LLA 12)& 0.590      & [OII]      &(  34.9  )& ASDL       \\       
A& 6C1113+3458& 2.33 & 0.52 & HEG?& $\phantom{>}18.29$ ~ (L    8)& 2.406?     & Ly$\alpha$?&   35.97  & REL        \\
 & 6C1123+3401& 3.40 &-0.27 & LEG?& $\phantom{>}17.92$ ~ (Eea  5)& 1.247?     &  CII]?     &   34.59  & REL        \\       
A& 6C1125+3745& 2.07 & 0.79 & Q   & $\phantom{>}17.46$ ~ (LLA 12)& 1.233      & [OII]      & $<35.15$ & REL        \\       
A& 6C1129+3710& 2.36 & 0.51 & HEG?& $\phantom{>}17.52$ ~ (Eea  5)& 1.060      & [OII]      &          & REL        \\       
A& 6C1130+3456& 3.20 & 0.57 & LEG & $\phantom{>}15.76$ ~ (LLA 12)& 0.512      & [OII]      &(  35.1)  & ALL        \\       
 & 6C1134+3656& 2.07 & 0.73 & HEG & $\phantom{>}19.40$ ~ (ER   5)& 2.125      & Ly$\alpha$ &   36.62  & REW$\dag$  \\

\hline

A& 6C1141+3525& 2.40 & 0.82 & HEG & $\phantom{>}18.72$ ~ (Eea  5)& 1.781      & Ly$\alpha$ &   37.00  & ASDL     \\
A& 6C1143+3703& 2.06 & 0.24 & LEG?& $\phantom{>}19.83$?~ (Eea  5)&      -     &           -&         -& REL      \\       
A& 6C1148+3638& 3.21 & 0.69 & LEG?& $\phantom{>}14.63$ ~ (LLA 12)& 0.141      & [OII]      &($>33.8$ )& S        \\       
A& 6C1148+3842& 3.83 & 0.64 & Q   &        -         & 1.303      &           -&          & WW       \\       
 & 6C1158+3433& 2.12 & 0.50 & LEG &        -         & 0.530?     & [OII]?     &   34.60  & REL      \\       
A& 6C1159+3651& 2.20 & 0.50 & LEG?& $\phantom{>}17.42$ ~ (L89  8)&      -     &           -&         -& REL      \\
A& 6C1204+3708& 3.92 & 0.61 & HEG & $\phantom{>}19.29$ ~ (Eea  5)& 1.779      & Ly$\alpha$ &   37.00  & REL      \\
A& 6C1204+3519& 3.43 & 0.54 & HEG & $\phantom{>}17.97$ ~ (Eea  5)& 1.376      & CIV        &   35.73  & REL      \\      
 & 6C1205+3912& 3.83 & 0.76 & LEG?&        -         & 0.243      & [OII]      &(  34.9  )& S        \\       
A& 6C1212+3805& 2.14 & 0.49 & LEG?& $\phantom{>}17.74$ ~ (Eea  5)& 0.947?     & [OII]?     &   35.20  & REL      \\       

\hline

 & 6C1213+3504& 2.39 & 0.02 & Q   &        -         & 0.857      & [OII]      &(  35.4  )& VT       \\       
A& 6C1217+3645& 2.40 & 0.74 & HEG?& $\phantom{>}17.27$ ~ (Eea  5)& 1.089?     & [OII]?     &   34.81  & REL      \\       
A& 6C1220+3723& 2.52 & 0.59 & Q   & $\phantom{>}15.30$ ~ (LLA 12)& 0.489      & [OIII]     &(  36.2  )& ALL      \\       
A& 6C1230+3459& 2.90 & 0.42 & HEG & $\phantom{>}18.40$ ~ (Eea  5)& 1.533      & [NeIV]     &   35.59  & ASDL     \\       
 & 6C1232+3942& 3.27 & 0.93 & HEG & $\phantom{>}17.80$ ~ (Ea  11)& 3.221      & Ly$\alpha$ &   36.73  & REW      \\
A& 6C1255+3700& 3.66 & 0.49 & Q   & $\phantom{>}15.59$ ~ (LLA 12)& 0.710      & [OIII]     &(  36.1  )& EH$\ddag$\\      
A& 6C1256+3648& 2.88 & 0.60 & HEG?& $\phantom{>}17.75$ ~ (Eea  5)& 1.127      & [OII]      &   35.37  & REL      \\       
 & 6C1257+3633& 2.40 & 0.84 & HEG & $\phantom{>}17.51$ ~ (Eea  5)& 1.003      & [OII]      &   36.58  & REL      \\
A& 6C1301+3812& 3.46 & 0.63 & HEG & $\phantom{>}14.66$ ~ (LLA 12)& 0.470*     & [OII]      &(  35.5  )& ALL      \\

\hline\hline
\end{tabular}
{\caption[Table 1]{{\bf (cont).}
}}
\end{center}
\end{table*}
\normalsize

\clearpage

\clearpage

\normalsize

\begin{figure*}
\begin{center}
\setlength{\unitlength}{1mm}
\begin{picture}(150,180)
\put(80,133){\includegraphics{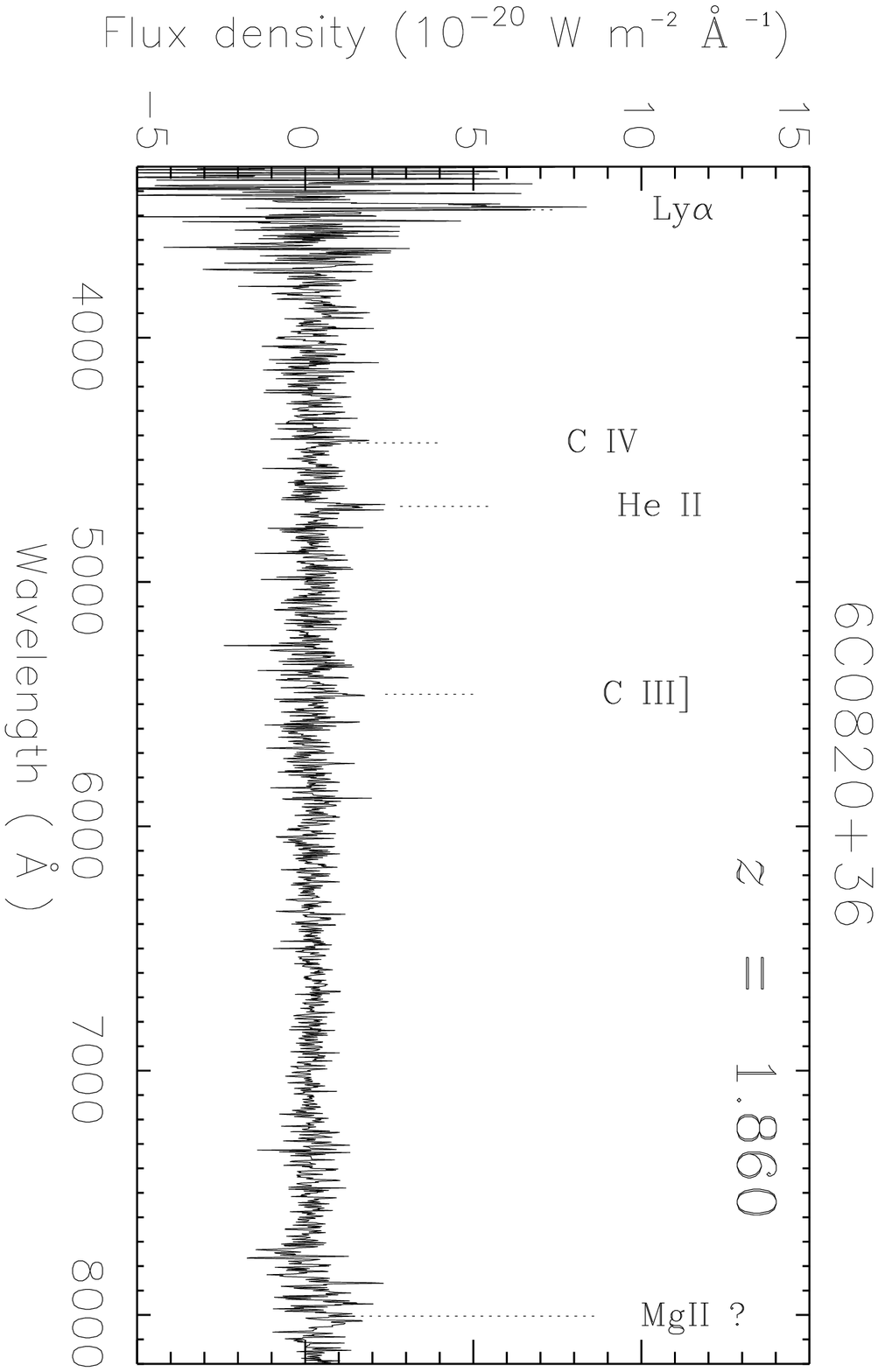}}
\put(80,111){\includegraphics{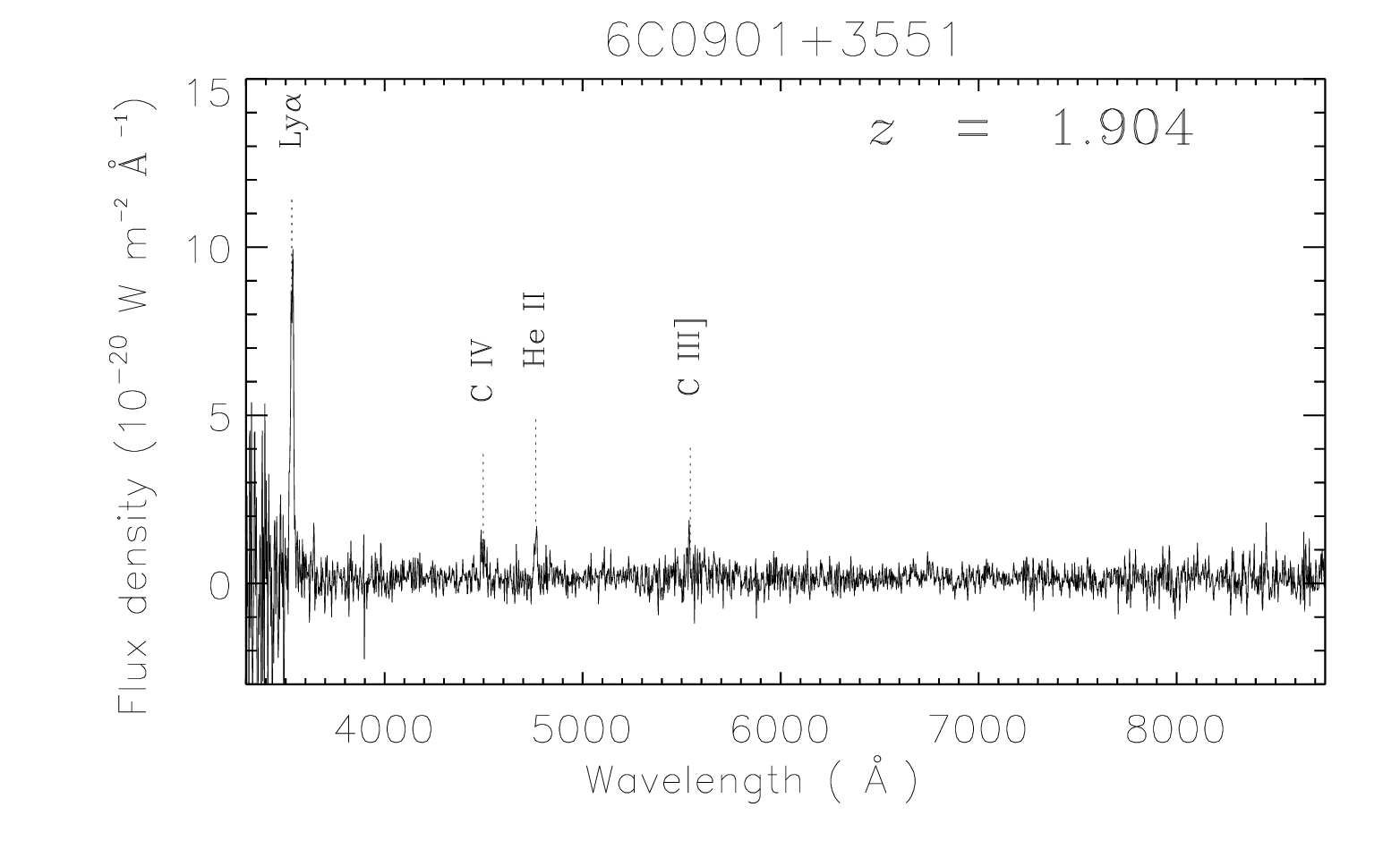}}
\put(80,78){\includegraphics{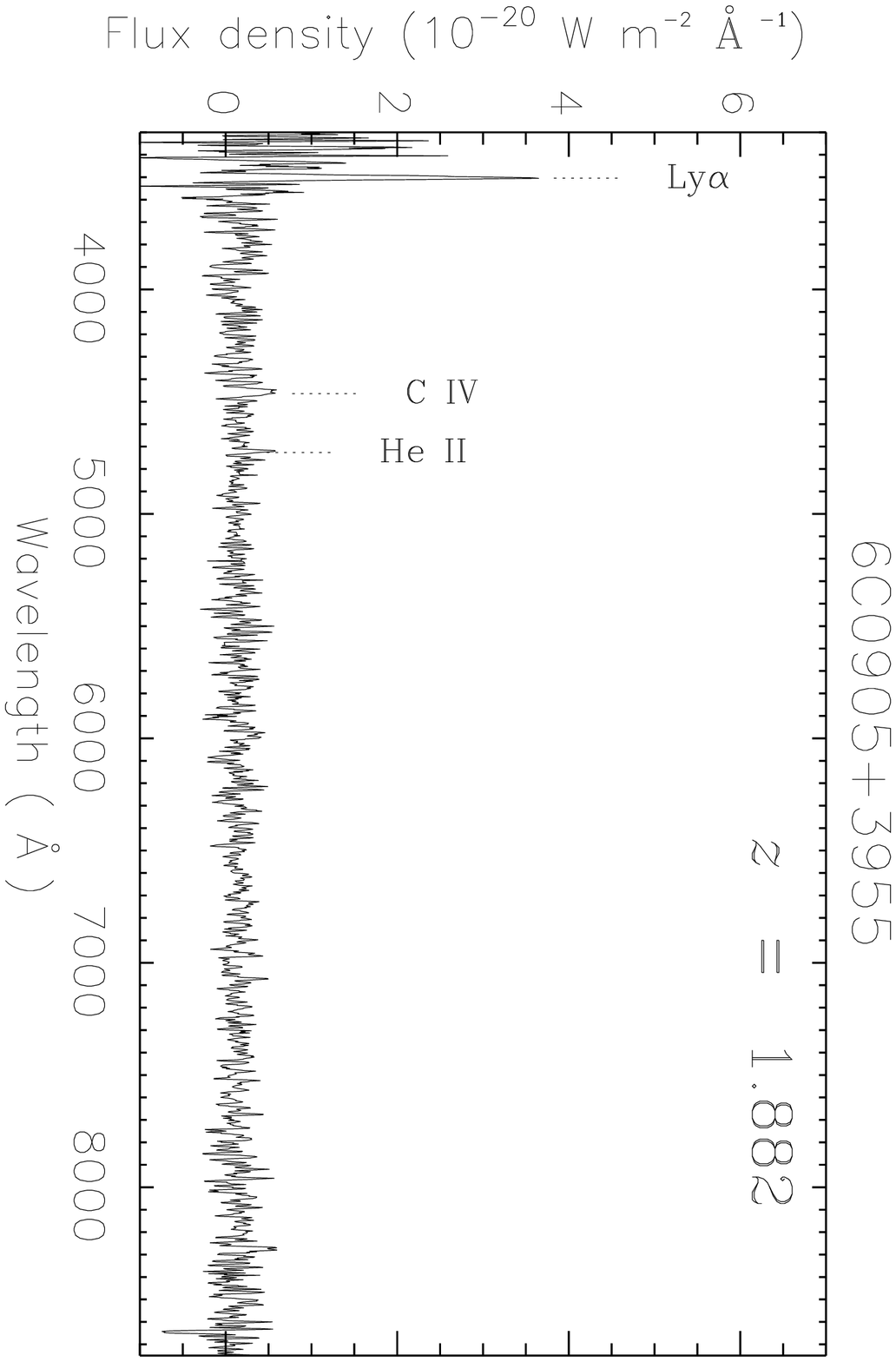}}
\put(163,78){\includegraphics{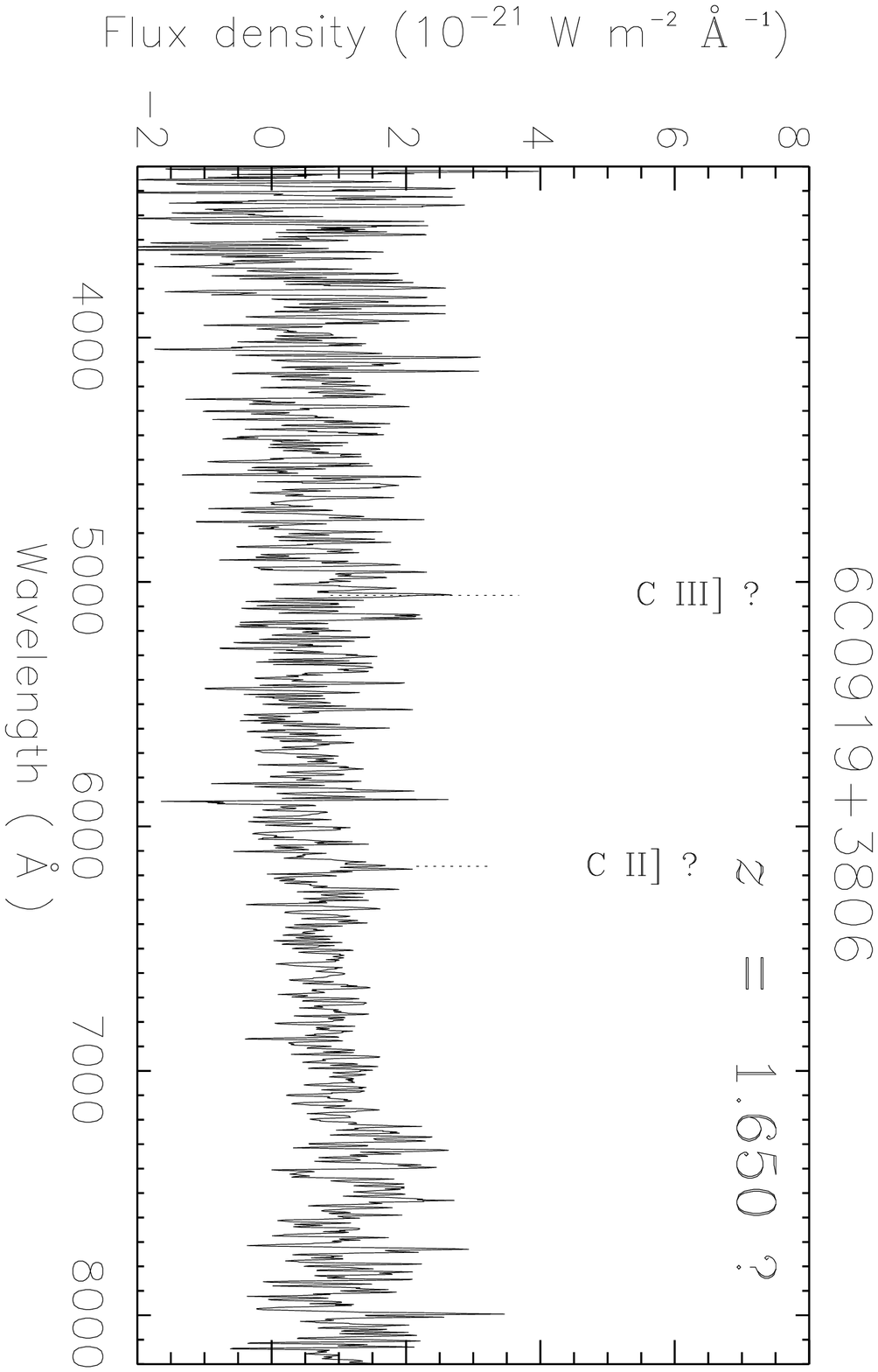}}
\put(80,23){\includegraphics{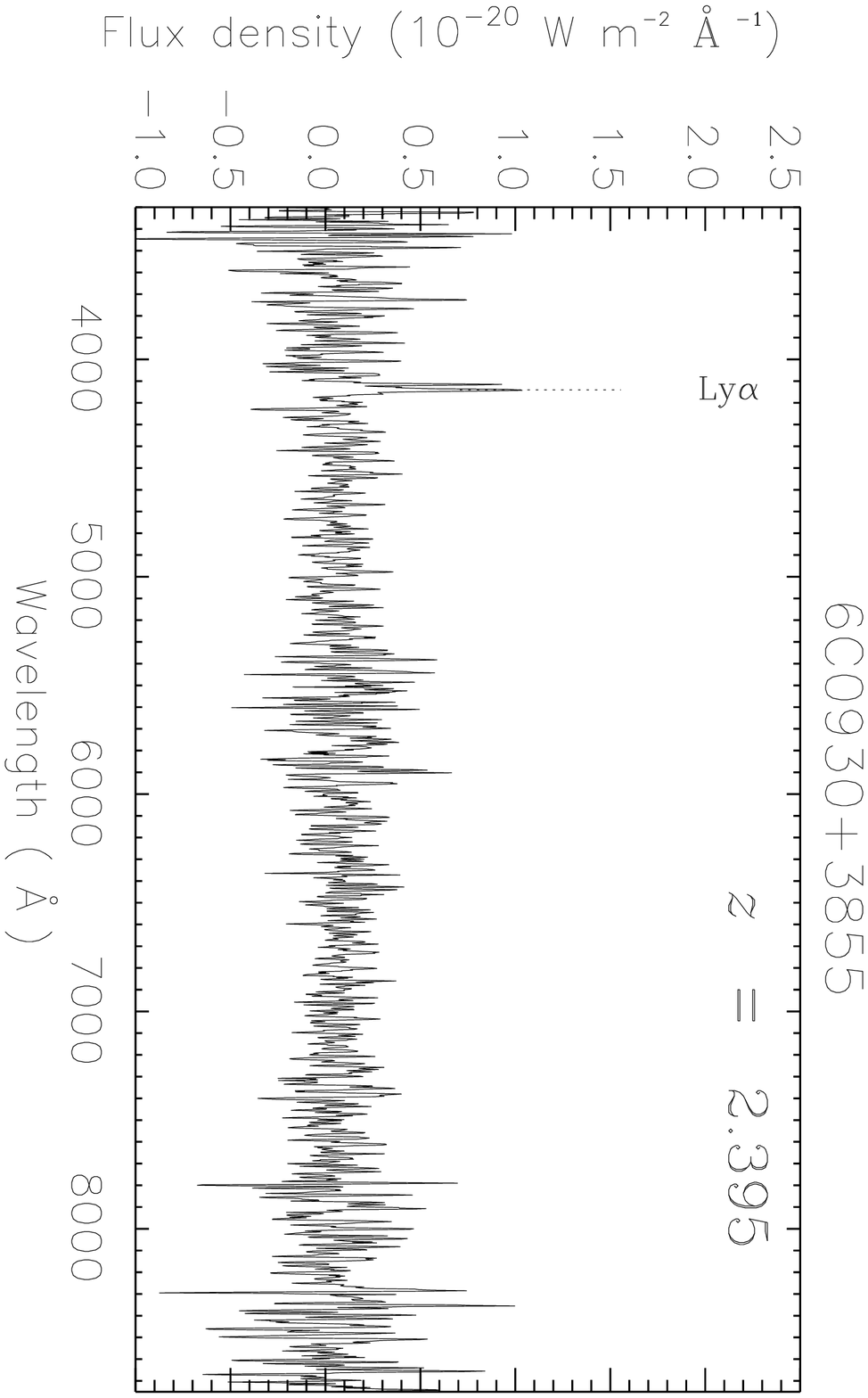}}
\put(163,23){\includegraphics{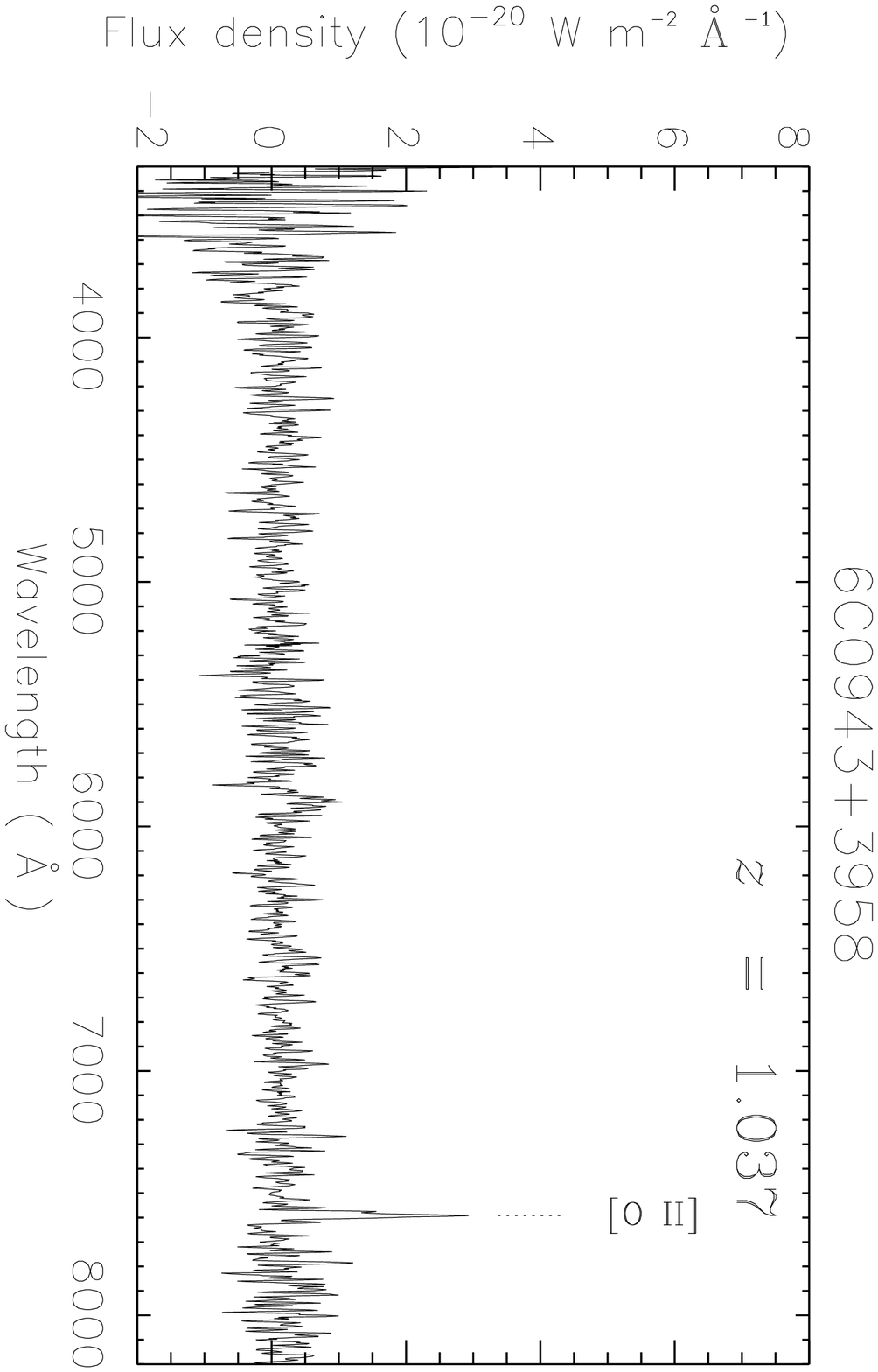}}
\put(80,-32){\includegraphics{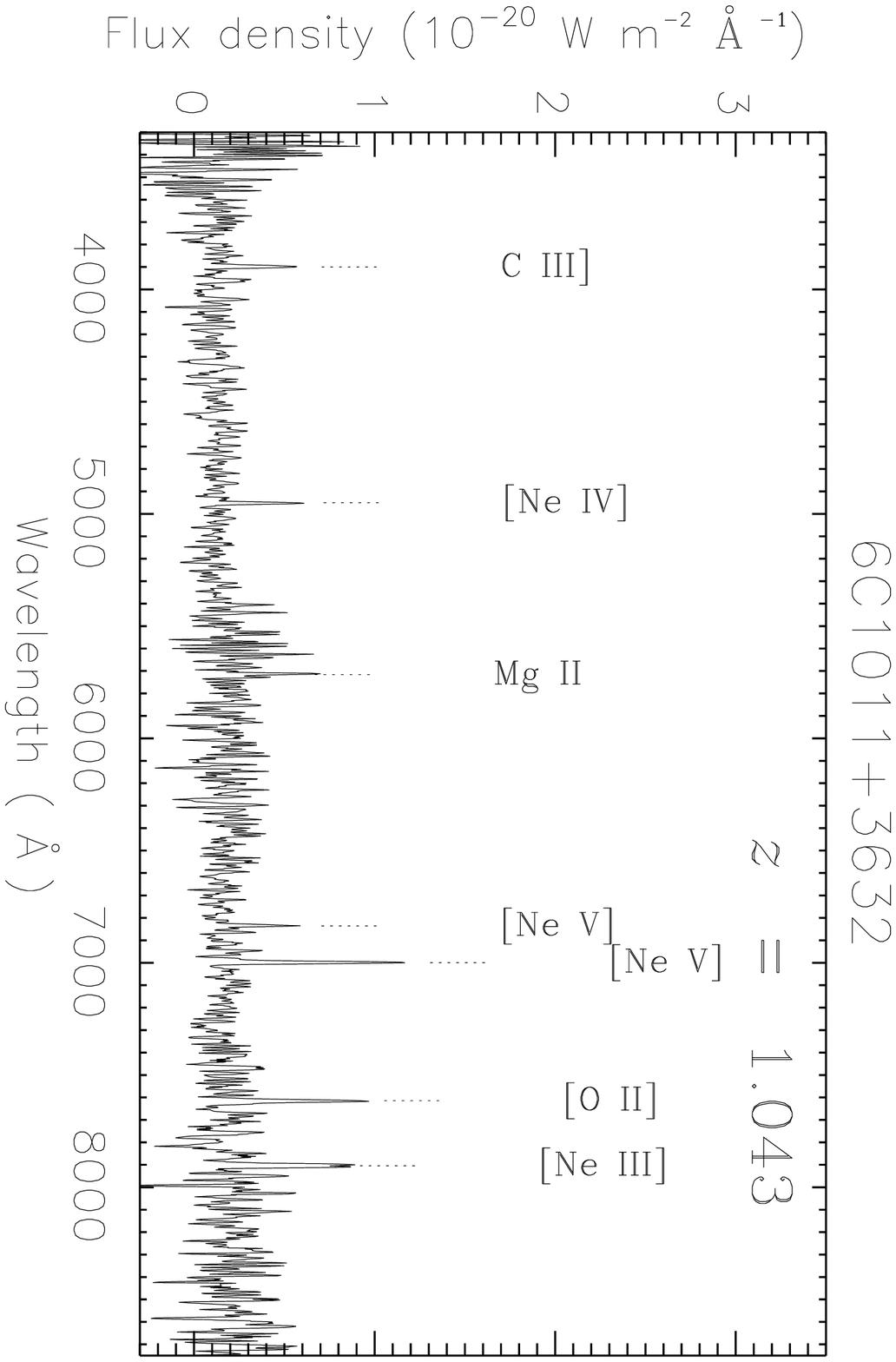}}
\put(163,-32){\includegraphics{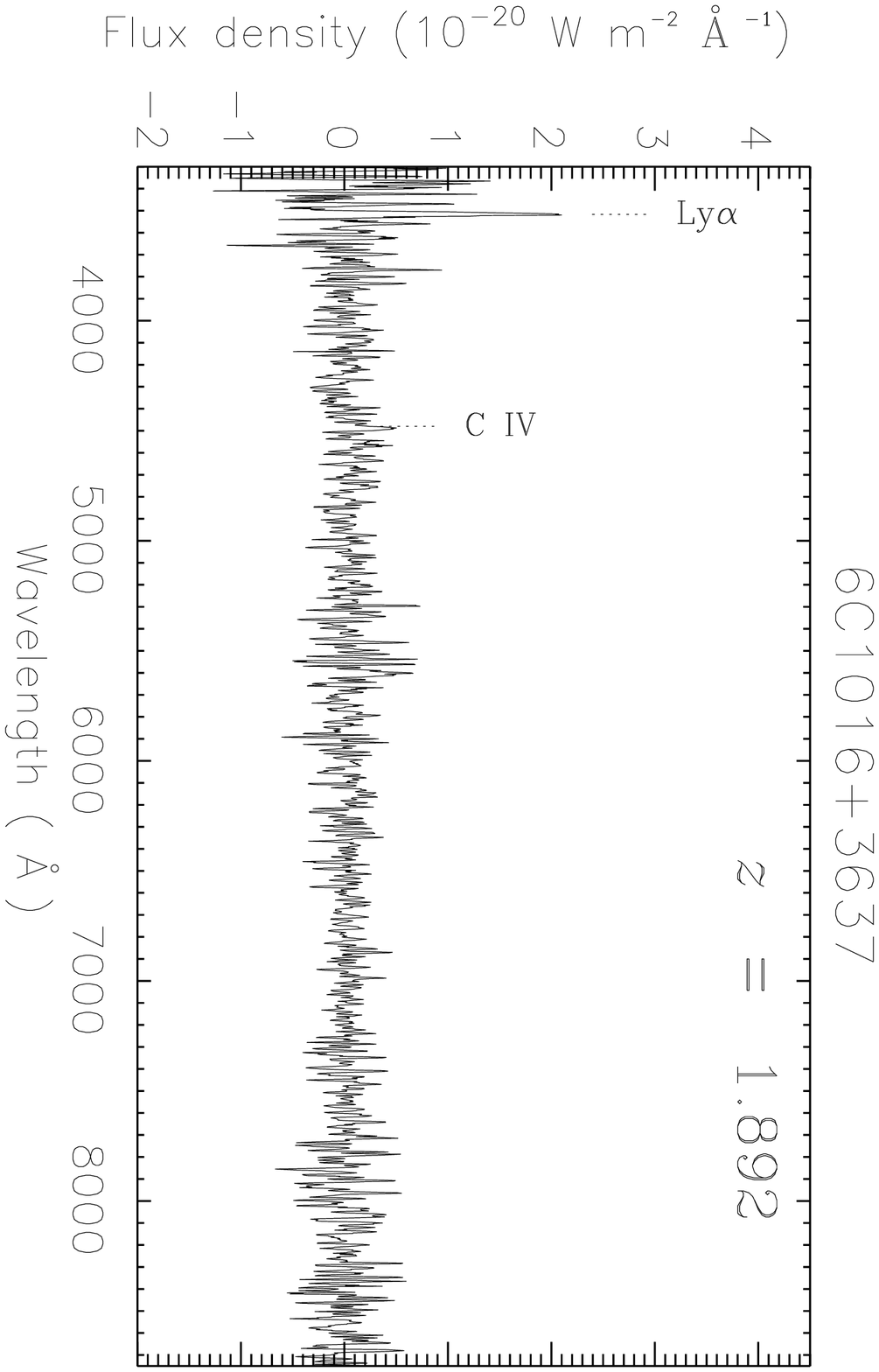}}
\end{picture}
\end{center}
\vspace{1.3in}
{\caption[junk]{\label{fig:spectra}
Spectra of the 6CE WHT targets with definite or possible spectral features.
The synthetic aperture used to extract 1D spectra from the 2D data
was typically defined by the full-width at 
zero-intensity of a cross-cut through the data, excepting the
cases of 6C1158+3433 and
6C1217+3645 for which a full-width at half-maximum
aperture was used.
}}
\end{figure*}

\addtocounter{figure}{-1}

\clearpage

\begin{figure*}
\begin{center}
\setlength{\unitlength}{1mm}
\begin{picture}(150,180)
\put(80,133){\includegraphics{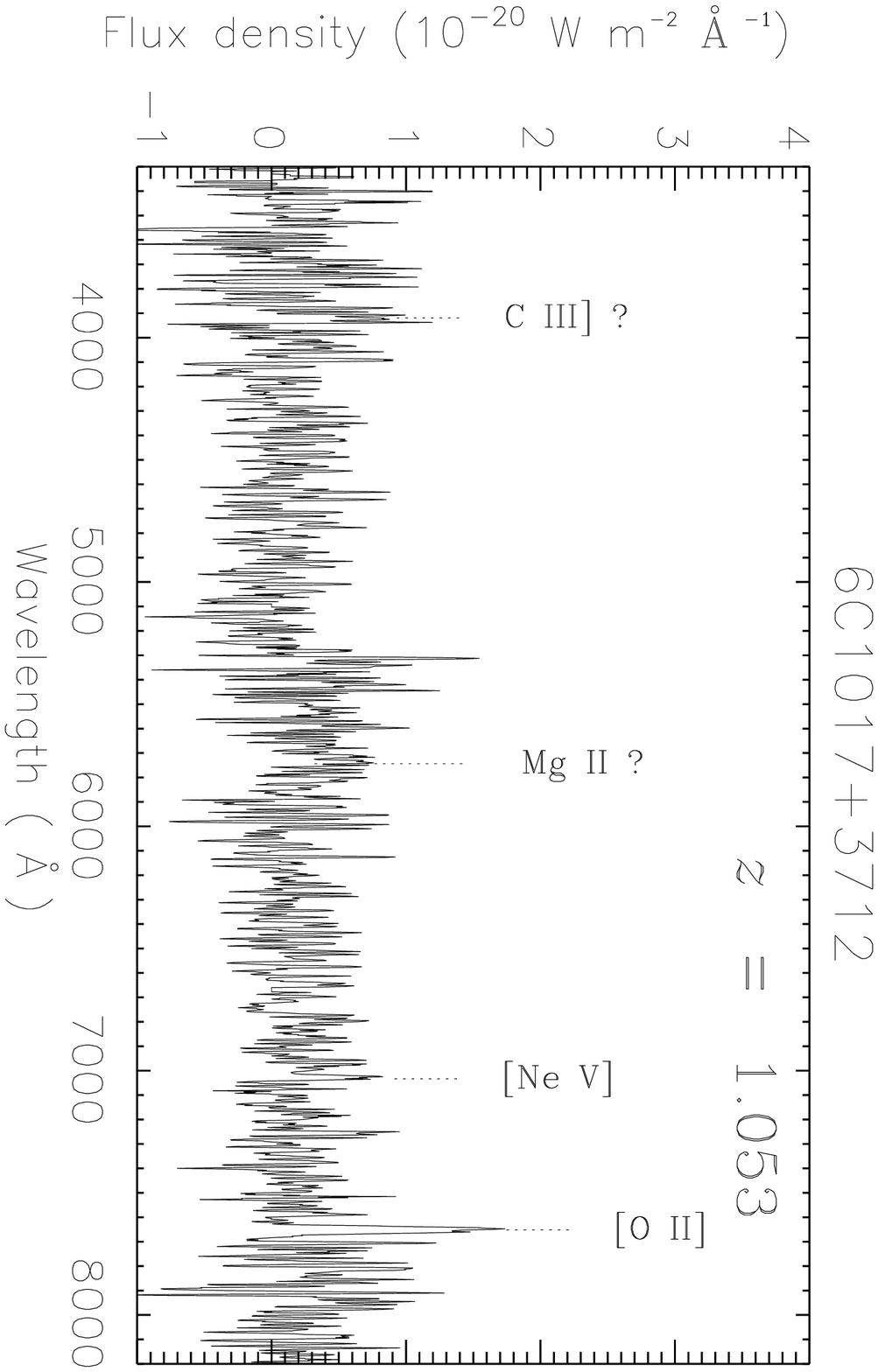}}
\put(163,133){\includegraphics{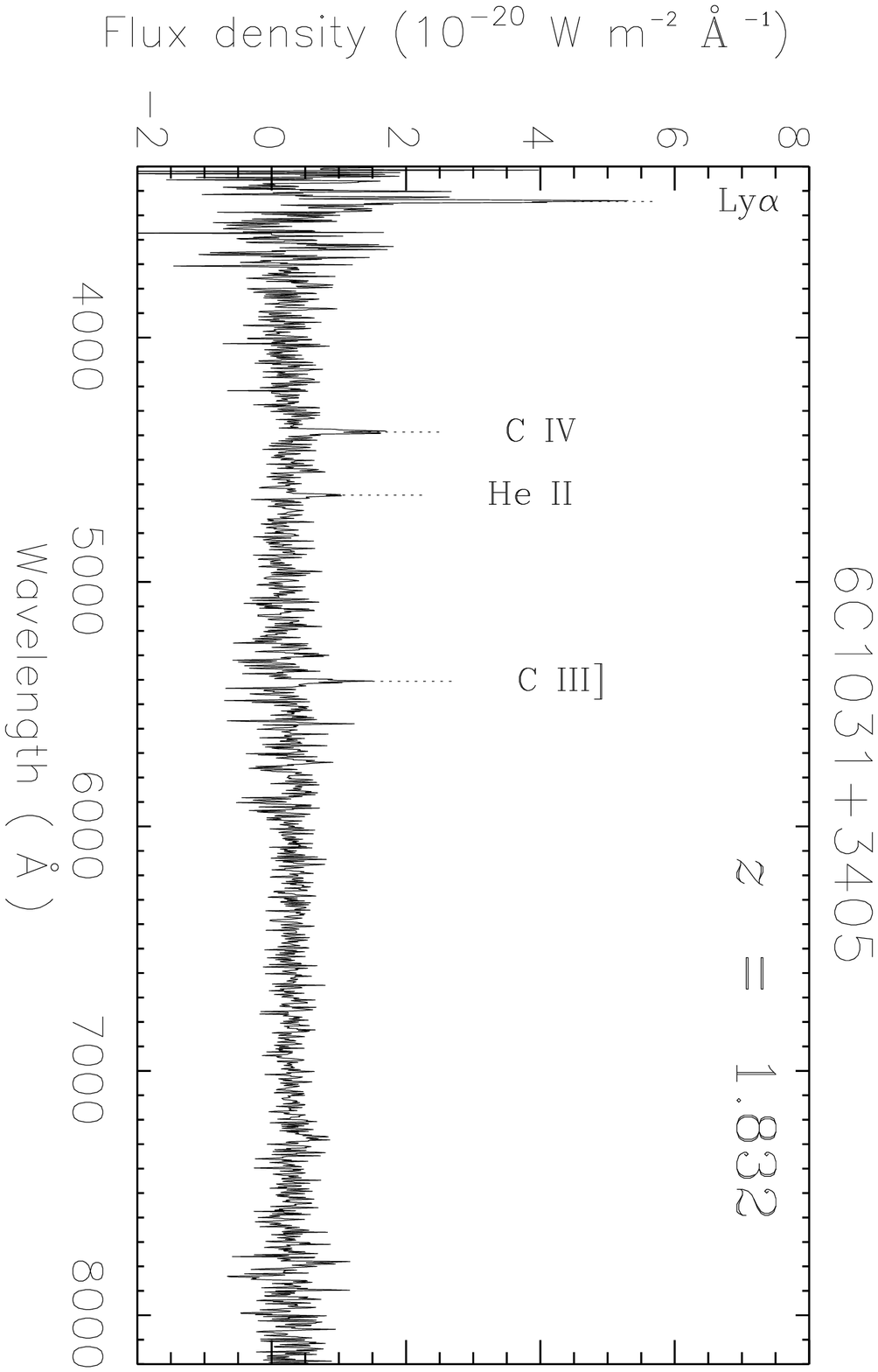}}
\put(80,78){\includegraphics{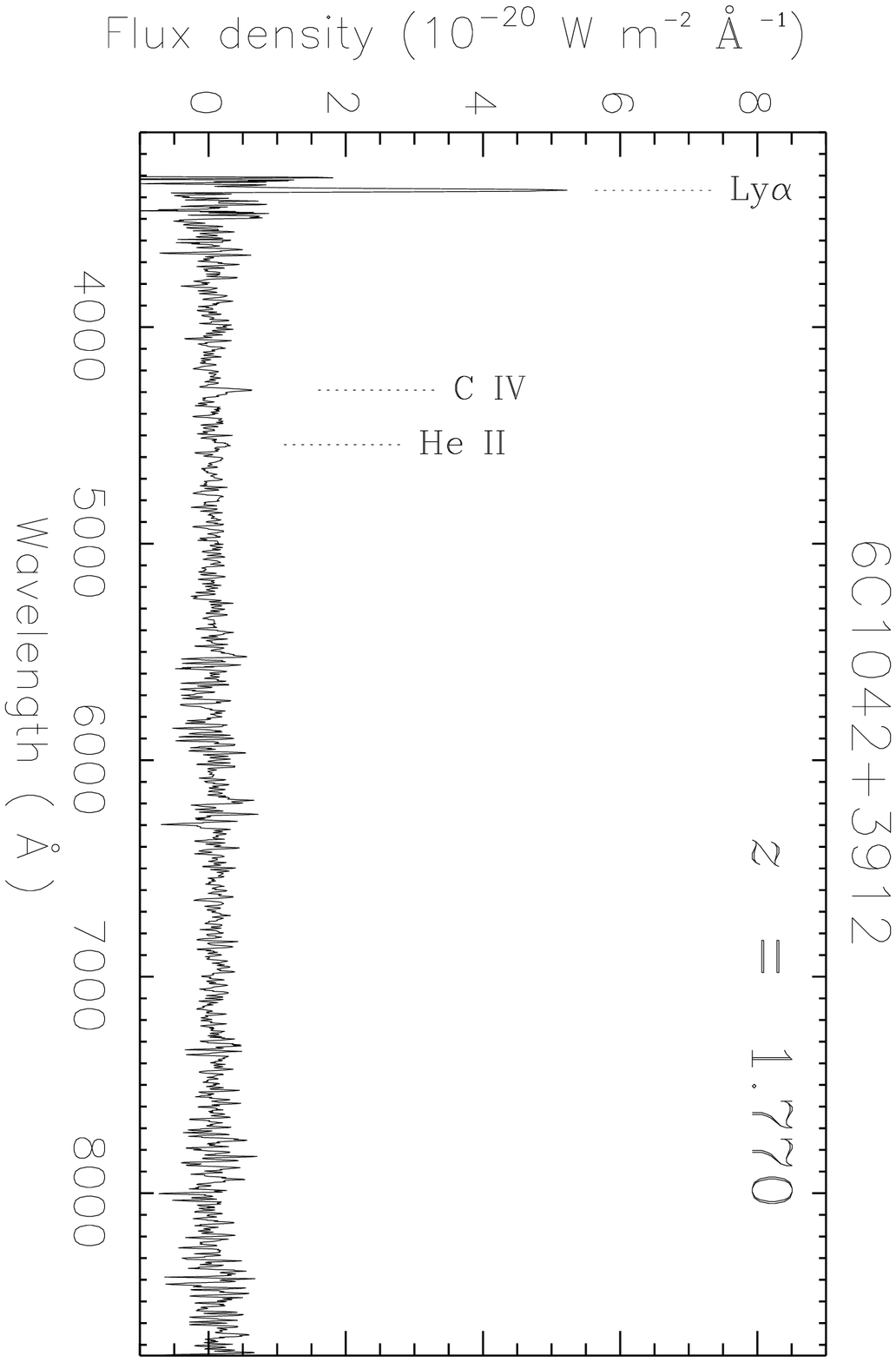}}
\put(163,78){\includegraphics{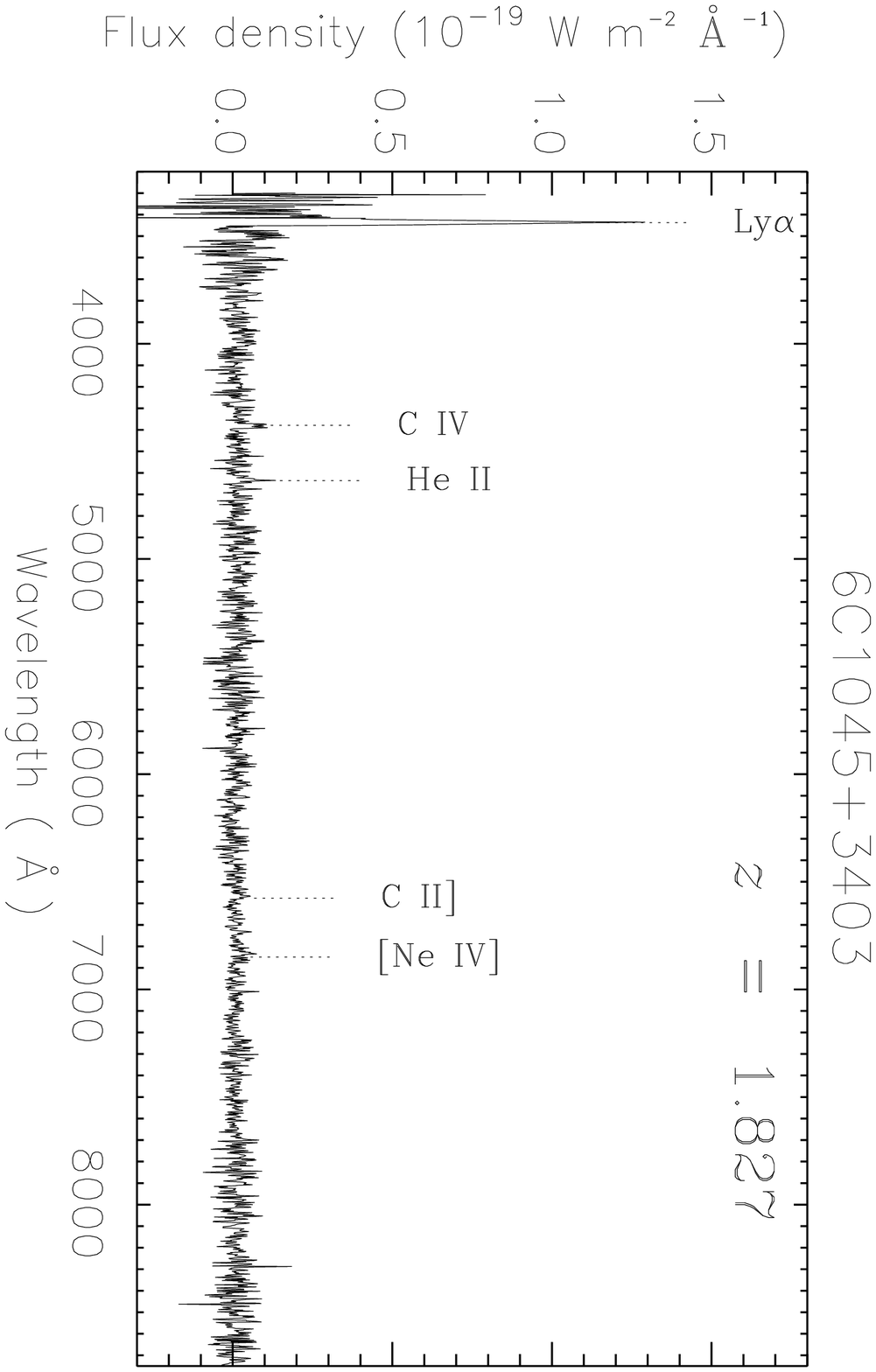}}
\put(80,23){\includegraphics{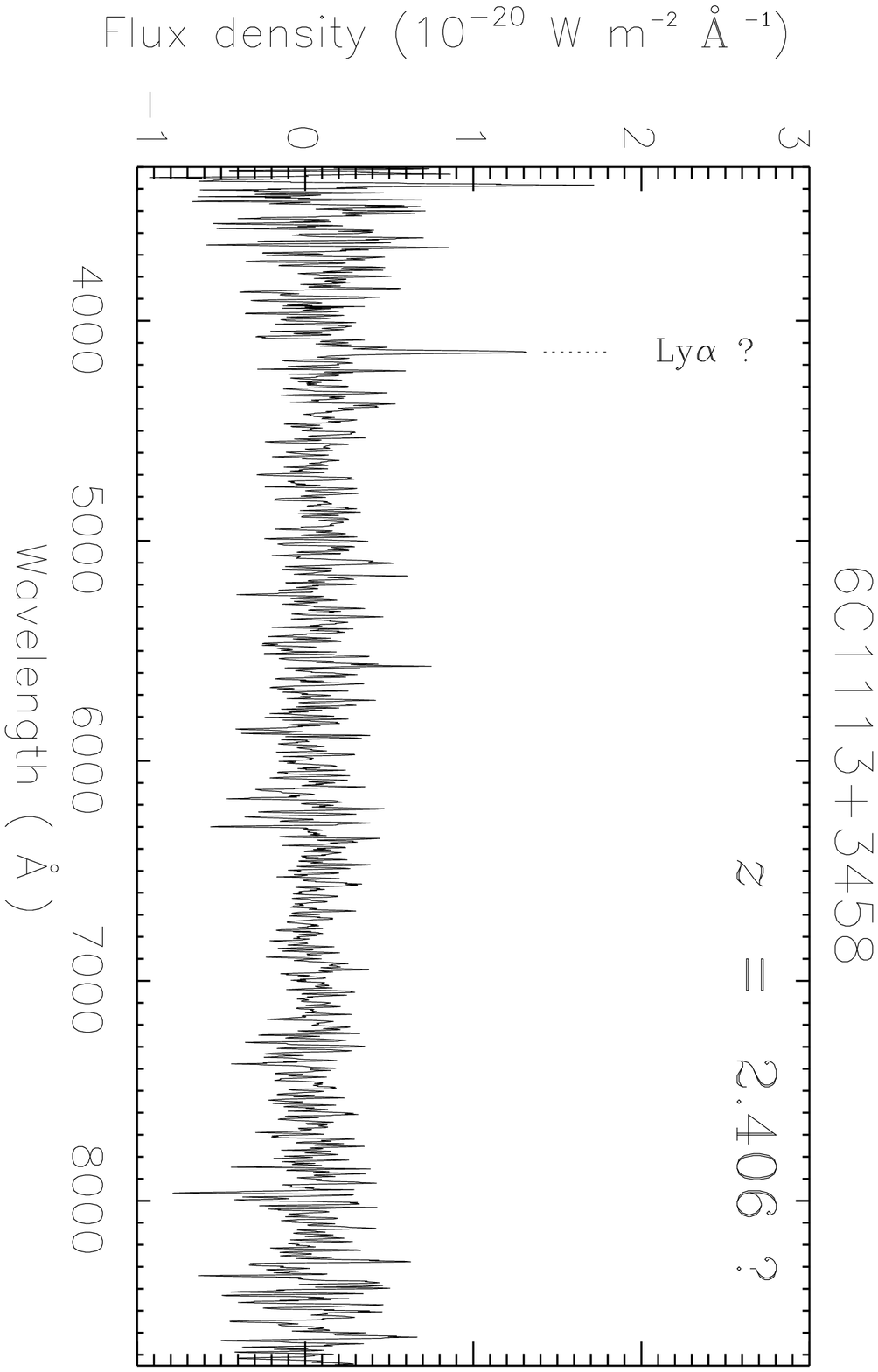}}
\put(163,23){\includegraphics{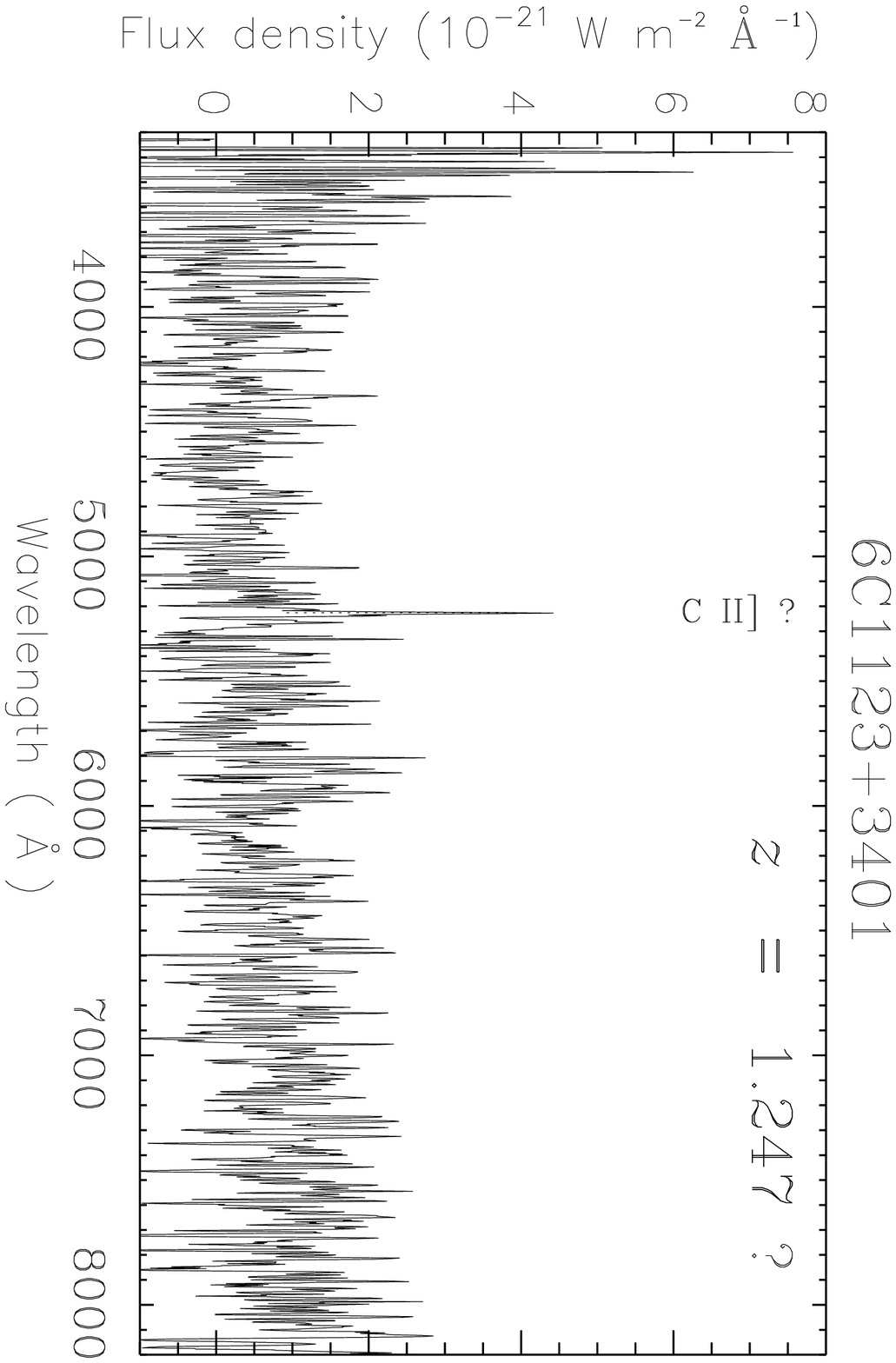}}
\put(80,-32){\includegraphics{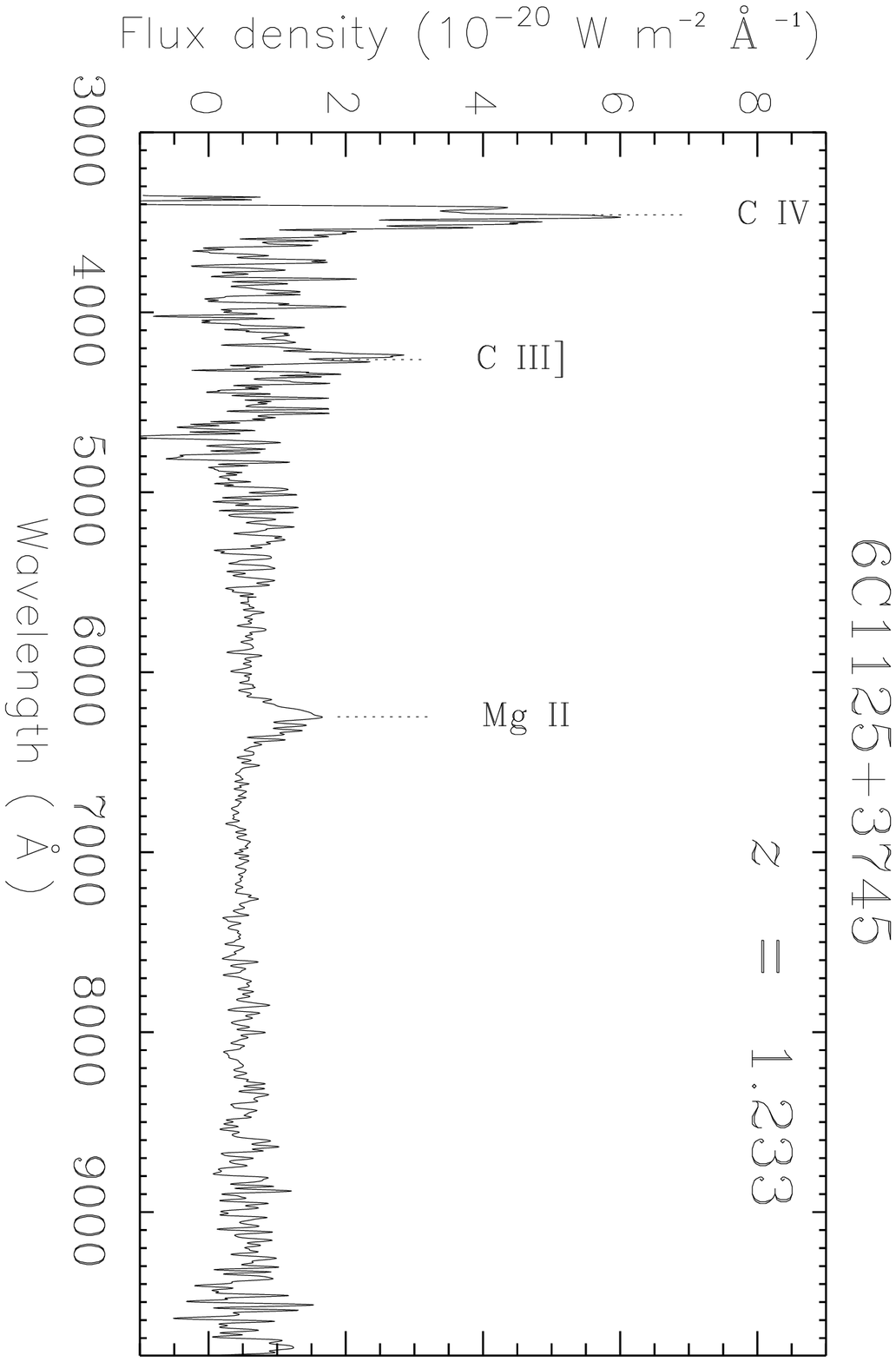}}
\put(163,-32){\includegraphics{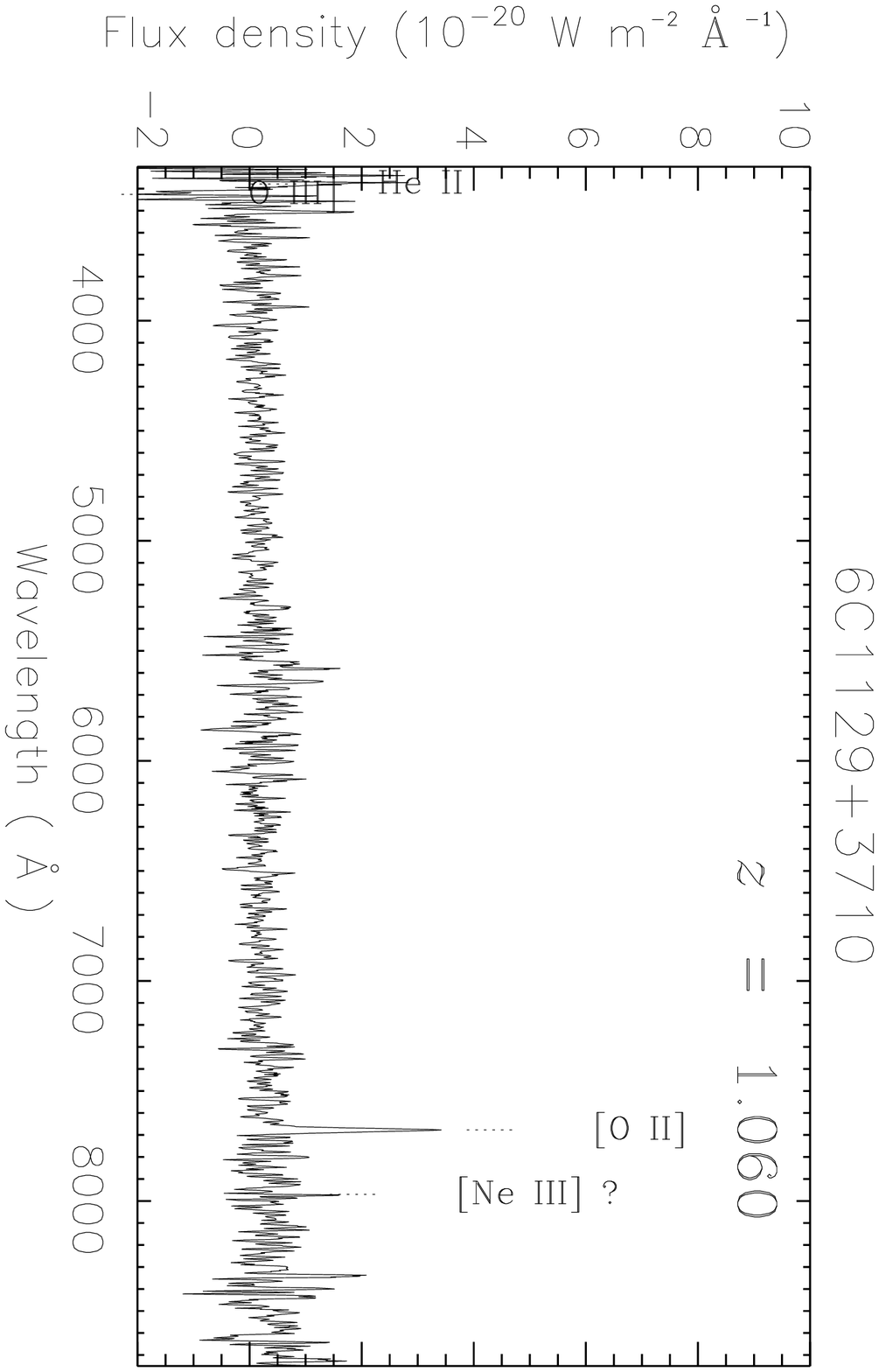}}
\end{picture}
\end{center}
\vspace{1.3in}
{\caption[junk]{{\bf (cont)}
}}
\end{figure*}

\addtocounter{figure}{-1}

\clearpage

\begin{figure*}
\begin{center}
\setlength{\unitlength}{1mm}
\begin{picture}(150,180)
\put(80,133){\includegraphics{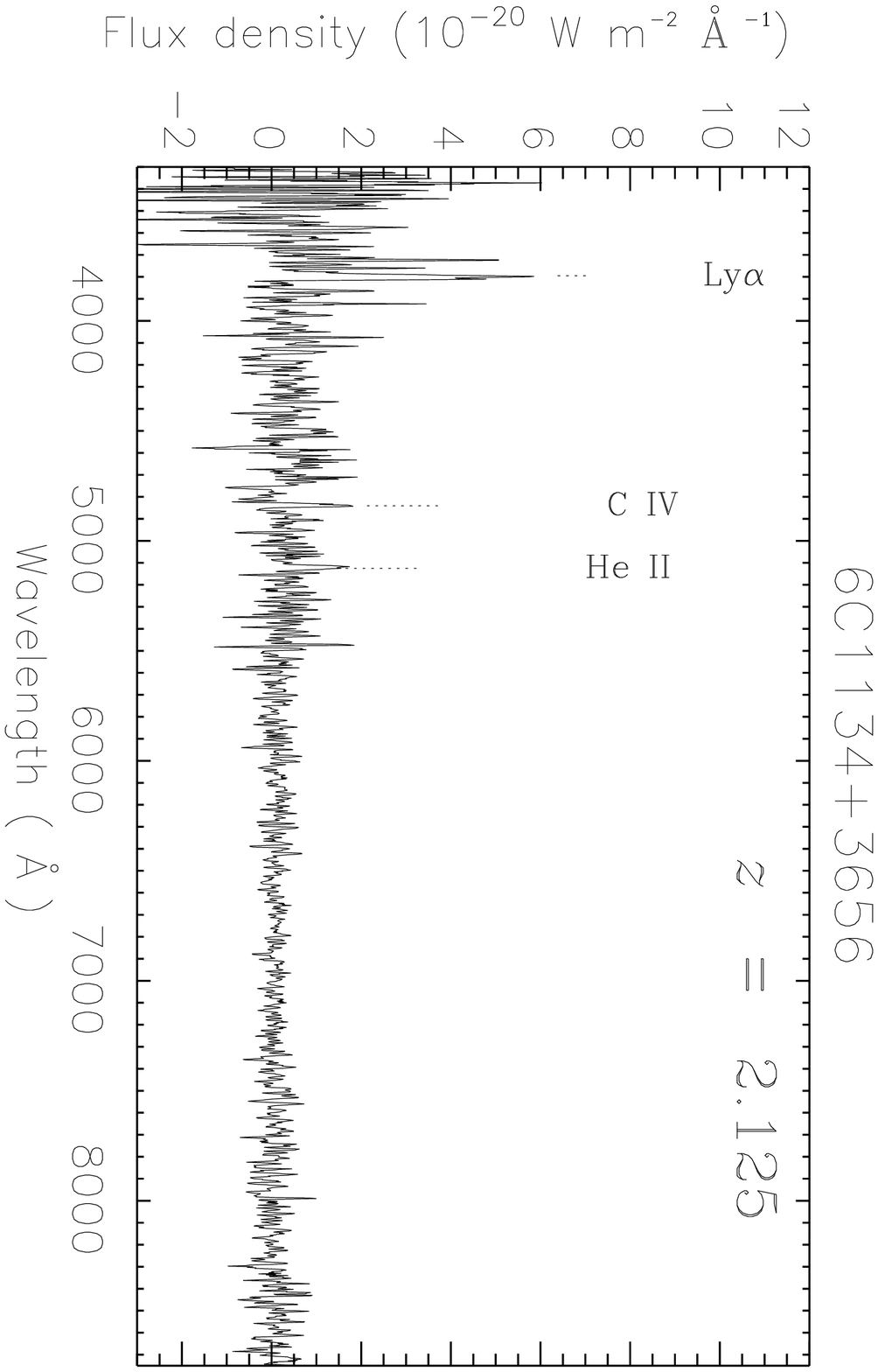}}
\put(163,133){\includegraphics{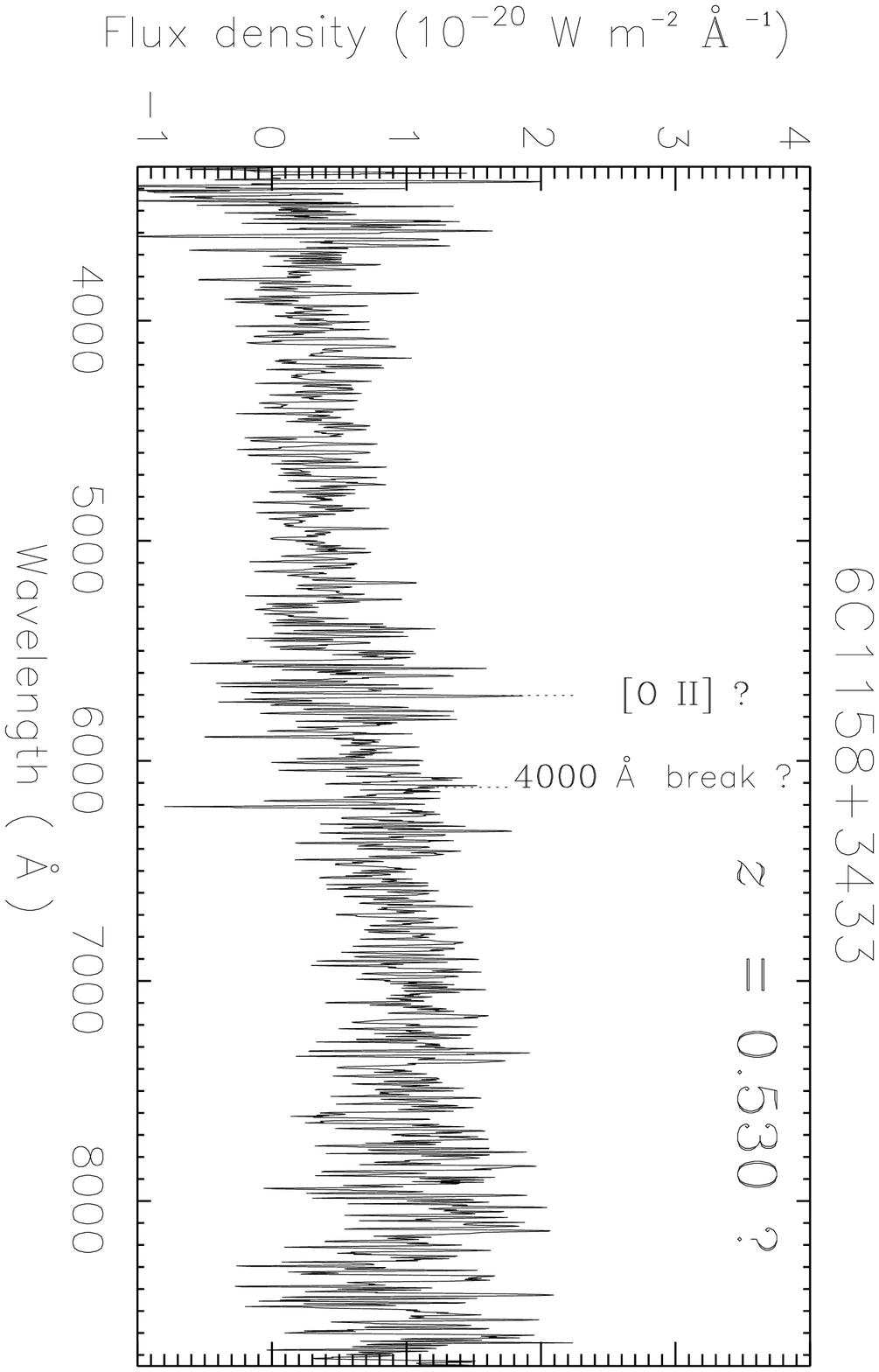}}
\put(80,78){\includegraphics{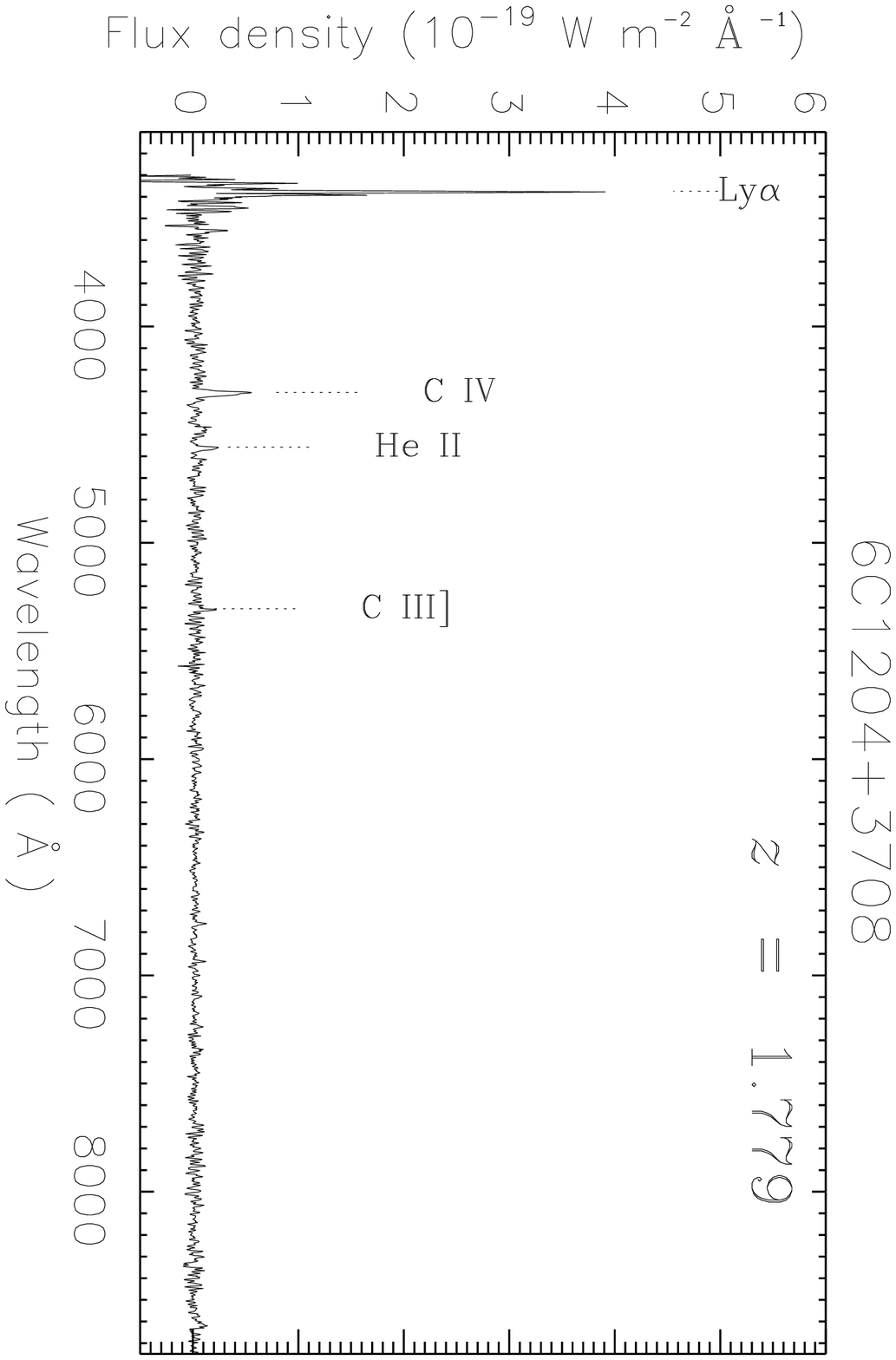}}
\put(163,78){\includegraphics{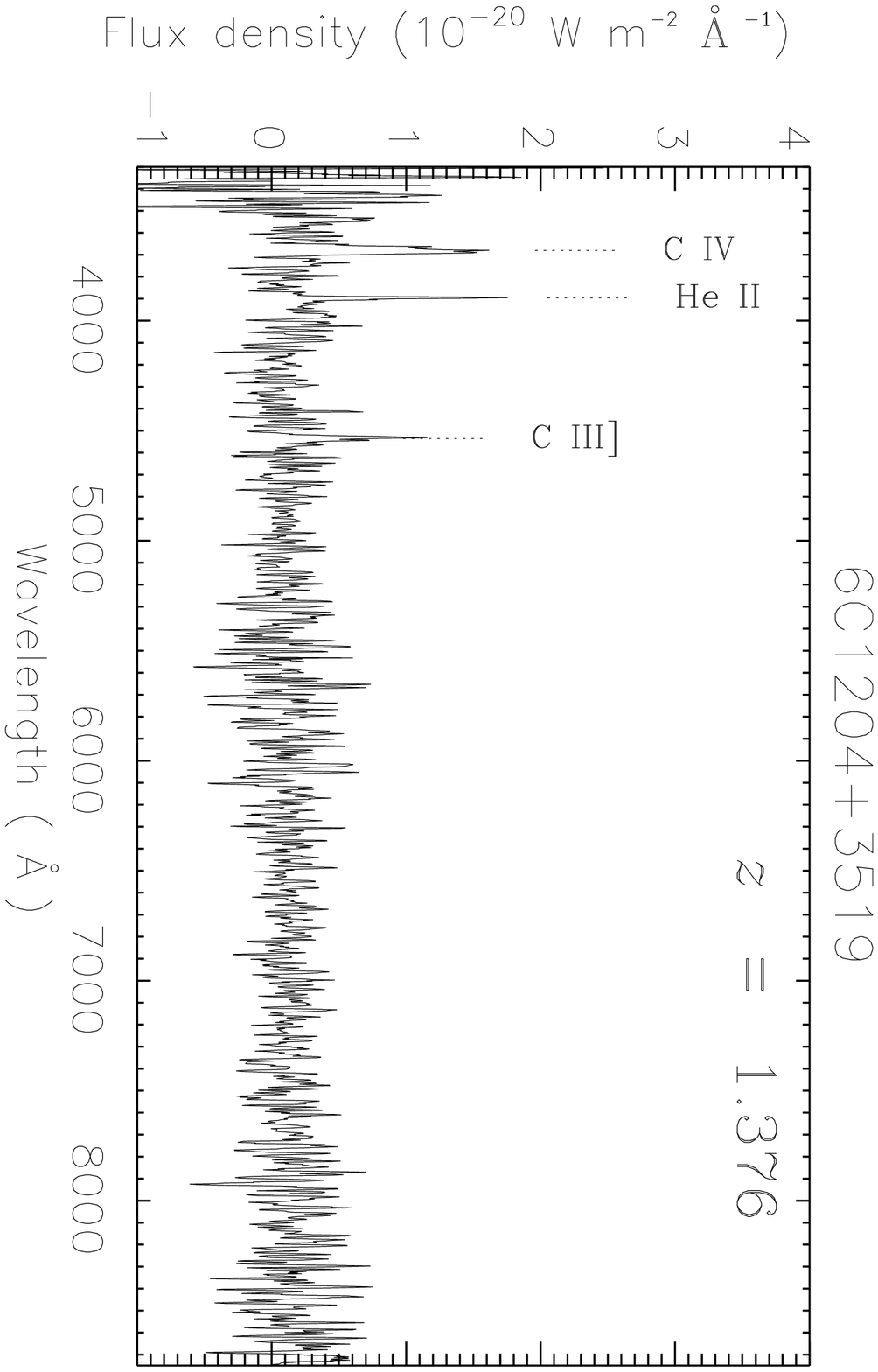}}
\put(80,23){\includegraphics{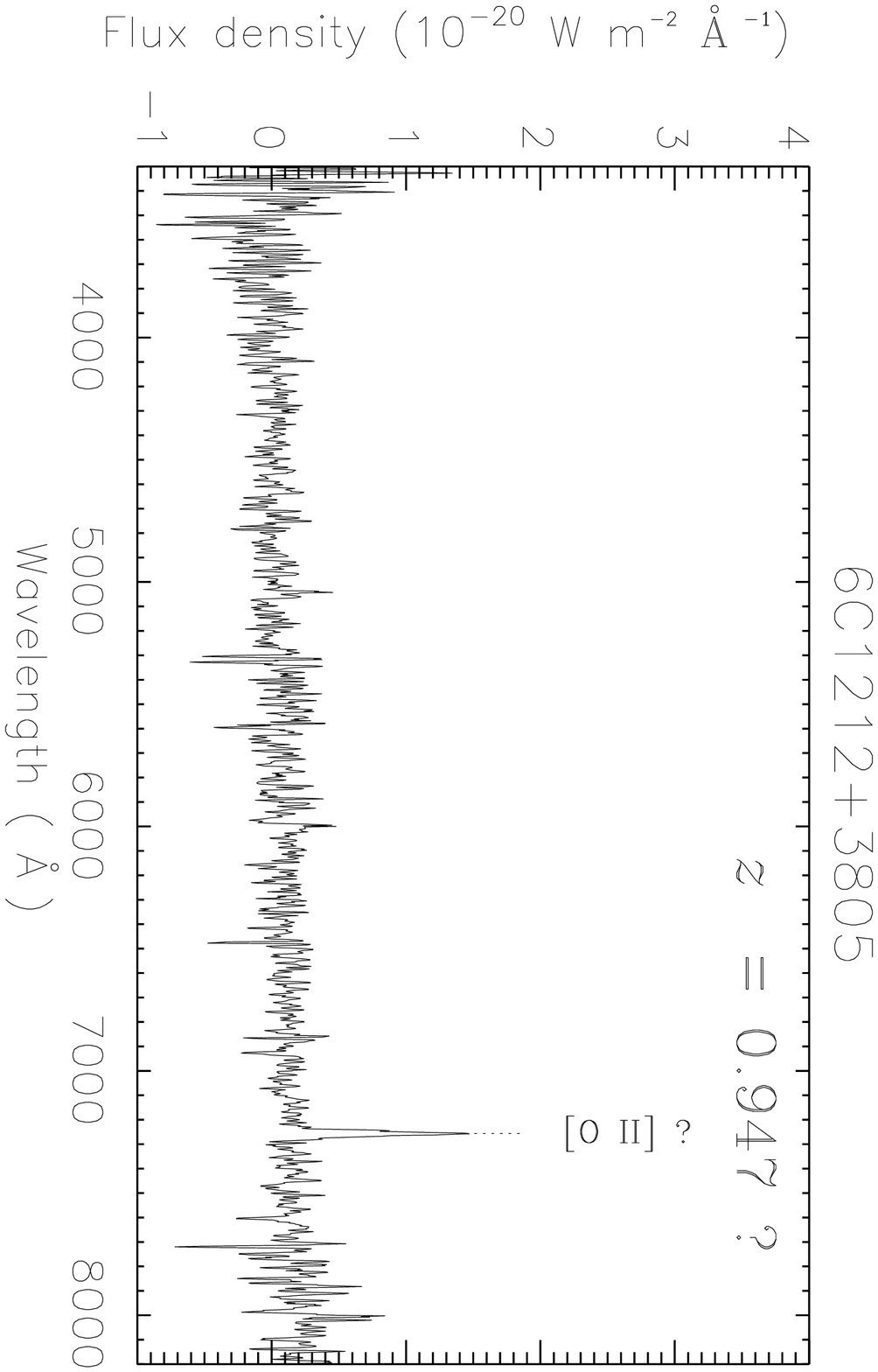}}
\put(163,23){\includegraphics{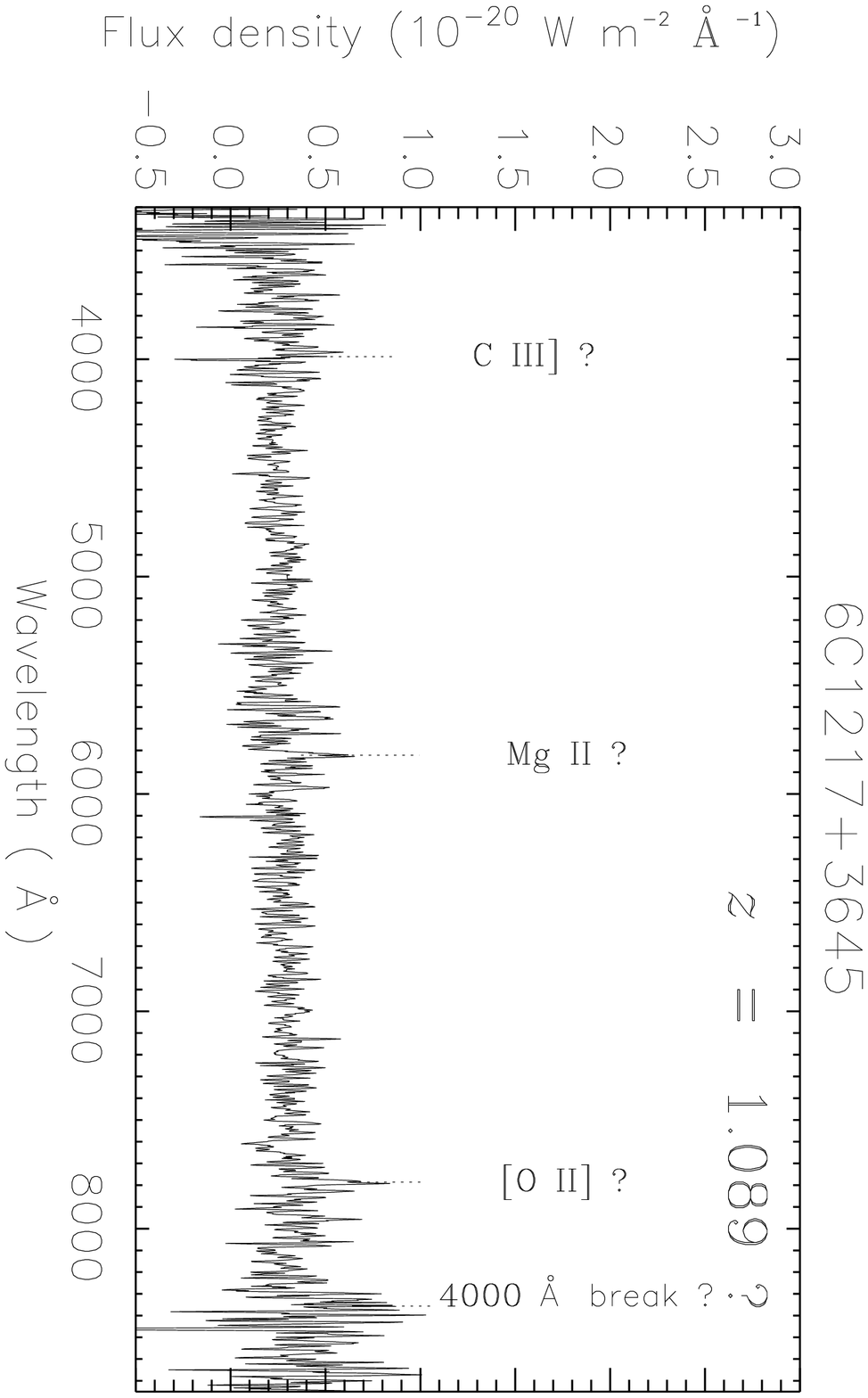}}
\put(80,-32){\includegraphics{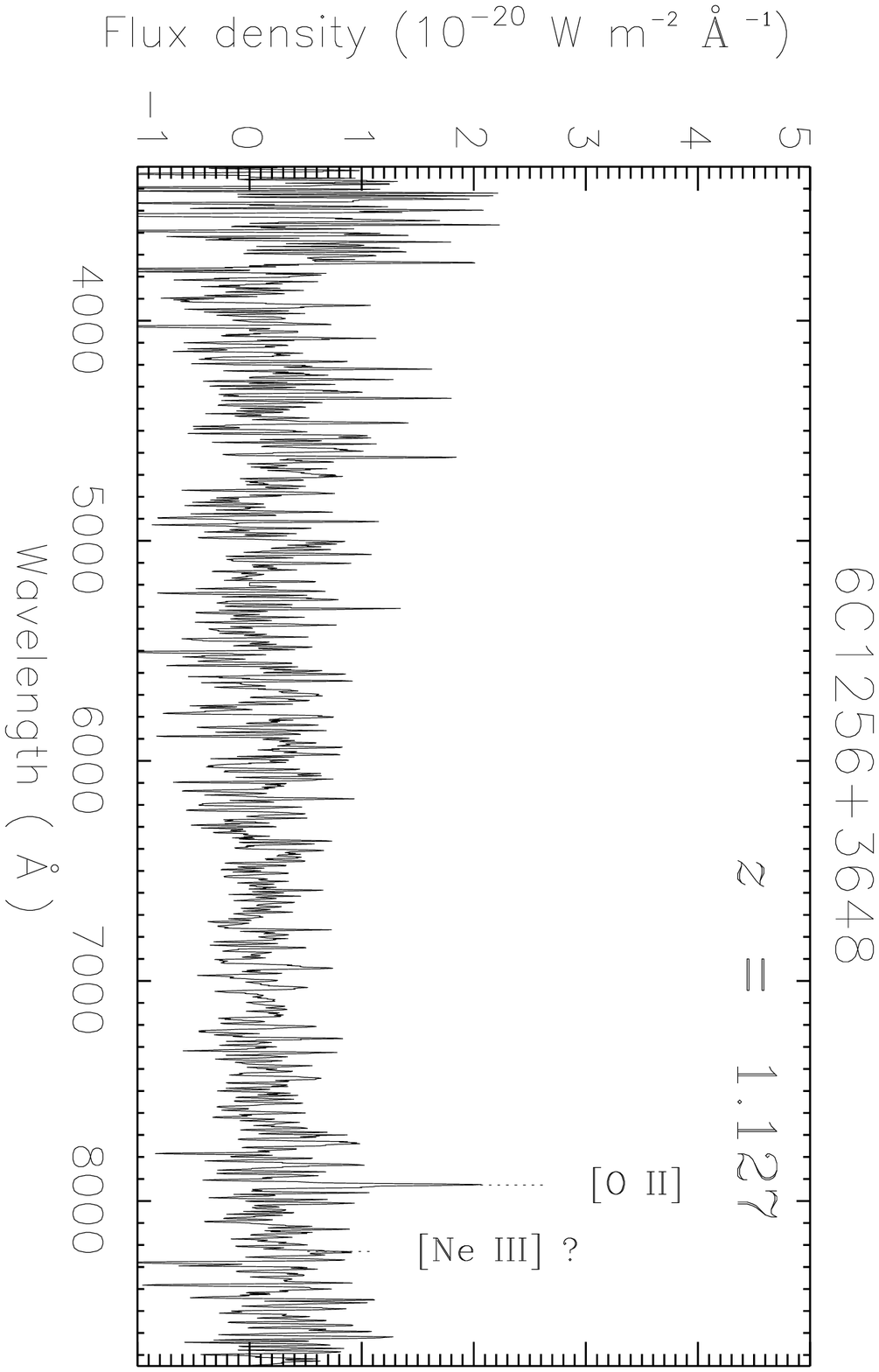}}
\put(163,-32){\includegraphics{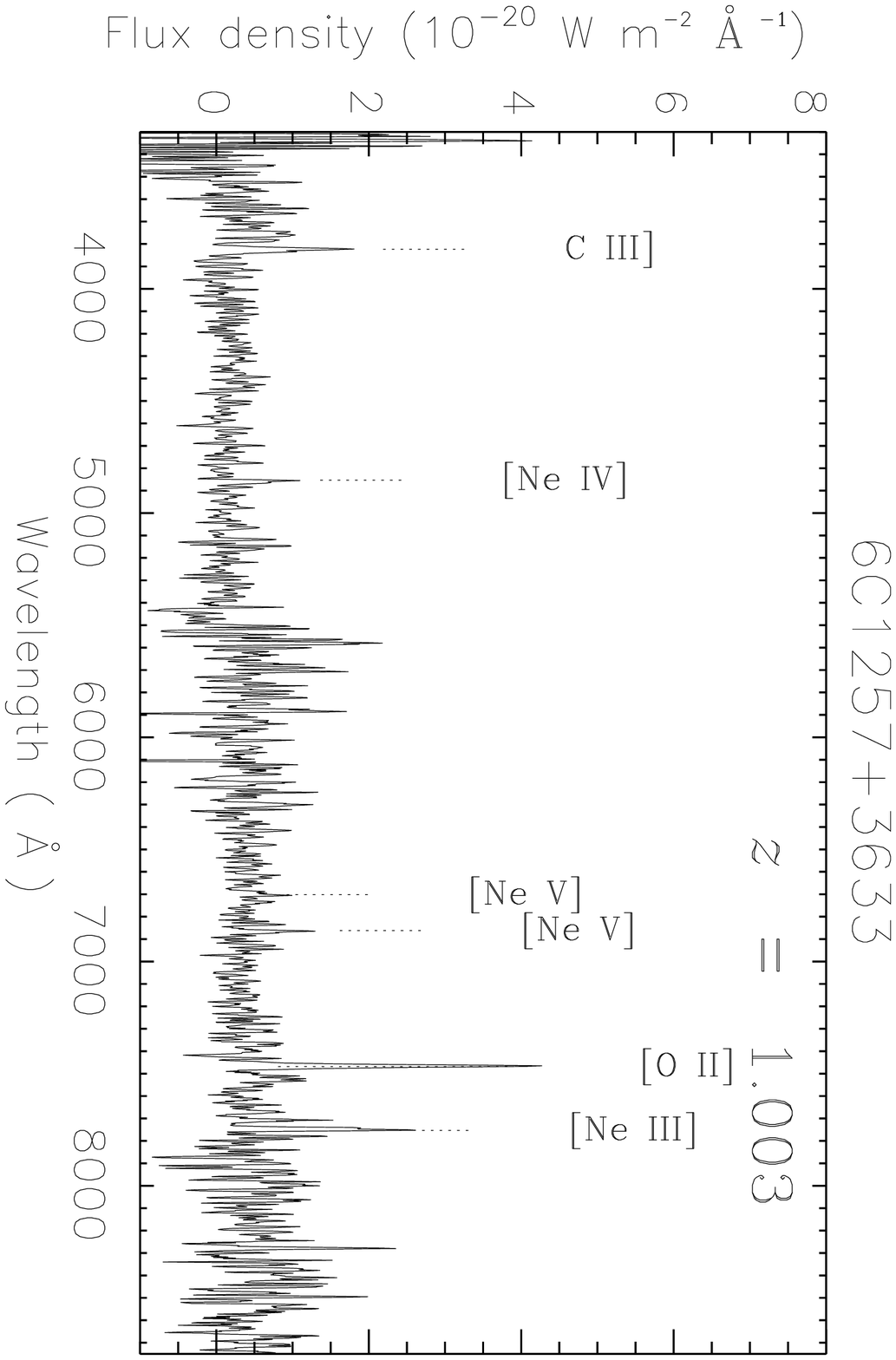}}
\end{picture}
\end{center}
\vspace{1.3in}
{\caption[junk]{{\bf (cont)}
}}
\end{figure*}

\clearpage

\normalsize

\subsection{Observations and spectroscopic redshifts}
\label{sec:spectra}

The optical spectra presented in this paper were mostly obtained
during the course of two WHT observing runs in Jan.\ 1994 and Jan.\ 1995 
using the ISIS spectrometer; some data taken with the FOS-II
spectrograph in Jan.\ 1992
are also presented (see Cotter, Rawlings \& Saunders 1996
for details of the observational-setup used in this run). A journal of all
these observations is presented in Table~2. 
All ISIS observations used the
R158R and R158B gratings with the beam split using the
5400-$\rm \AA$ dichroic; the blue-arm detector was the
TEK1 chip on both runs, the red-arm detector was the EEV3 chip in
1994 and the TEK2 chip in 1995. These setups produced continuous
spectral coverage from the blue atmospheric cut-off to
$\approx 8750 ~ \rm \AA$ in 1994, and to $\approx 8200 ~ \rm \AA$ in 1995.
The pixel scales for the TEK and EEV
chips were $0.358$ and $0.336$ arcsec pixel$^{-1}$ respectively.

Observations and data analysis 
followed the standard methods described by Lacy \& Rawlings (1994).
We offset from nearby bright stars
to either optical or radio positions, choosing the position angle
(PA) of the slit either on the basis of encompassing companion
optical objects or to coincide with the radio axis.
All the targets were observed below an air mass of 1.2, and all bar two
were observed below an air mass of 1.1. 
All the data presented here were taken in photometric
conditions with the seeing close to 1 arcsec. In nearly all cases
spectral features, emission lines and/or spectral breaks, were detected
and the reduced 1D spectra are presented in Fig.~\ref{fig:spectra}.
For cosmetic purposes, most of the spectra have been smoothed by replacing the value of each spectral bin by the average of the values in the bin and its two
neighbours. All measurements from the spectra were made prior to this 
process, and the results of these measurements are tabulated in
Table~\ref{tab:measurements}.

\subsubsection{Notes on the 6CE WHT targets}
\label{sec:obj_by_obj}

\begin{description}

\item[{\bf 6C0820+36}] We obtain 
an unequivocal redshift from the spatially-extended emission
lines seen in Fig.~\ref{fig:spectra}: the
alignment of the slit with the radio axis meant that we missed
the companion object to the south-west 
of the radio galaxy in the image of Eales et al.\ (1997).

\item[{\bf 6C0901+3551}] Our WHT spectrum confirms the tentative
redshift obtained by Rawlings, Eales \& Warren (1990).

\item[{\bf 6C0905+3955}] Our WHT spectroscopy of this giant radio galaxy
has been reported previously by Law-Green et al. (1995b):
the Ly$\alpha$ flux tabulated in Table~\ref{tab:measurements} 
is significantly higher than the value given in Table 2 of 
Law-Green et al.\ because it is evaluated from the total-light spectrum.
The Ly$\alpha$ emission extends from the continuum-emitting galaxy
towards the shorter western radio lobe with a secondary peak approximately midway 
between the radio galaxy and a companion object visible in the 
near-IR image of Law-Green et al.\

\item[{\bf 6C0919+3806}] Our WHT spectrum does not yield
an unequivocal redshift for this object although
two plausible emission features in the spectrum are consistent with 
$z=1.650$: the redshift given in 
Table 1 of Eales \& Rawlings (1996) is almost certainly 
incorrect. Continuum is detected from $3800 ~ \rm \AA$, and it remains 
roughly flat in $F_{\lambda}$ until $\approx 6100 ~ \rm \AA$, and then
rises (roughly $\propto \lambda^{3}$) 
into the red: the lack of a detected Lyman-limit cut-off means the redshift is
less than about $3.2$. We note broad similarities between this spectrum
and those of the old, red radio galaxies
studied by Dunlop (1999), and by identifying the continuum 
inflexion with a rest-frame wavelength $\approx 2600 ~ \rm \AA$, we obtain
a redshift estimate $z \sim 1.3$.

\item[{\bf 6C0930+3855}] Our WHT spectrum contains a single
doubled-peaked line which, since it matches well with lines
detected in the near-IR by Eales \& Rawlings (1996),
is securely identified with Ly$\alpha$. 
The cause of the double-peaked structure is most probably HI-absorption
in the host galaxy.

\item[{\bf 6C0943+3958}] Our WHT spectrum contains a
single bright emission line, but recent deeper spectroscopy in the
optical (P. Best \& S. Rawlings, {\it priv. comm.}) has confirmed the redshift
obtained if this feature is associated with [OII]3727.
The location of the strong line (and weak continuum) emission 
along the slit confirms that the radio galaxy lies much closer to the
eastern than the western radio hotspot in agreement with the 
near-IR position given by Eales et al. (1997).

\item[{\bf 6C1011+3632}] The spectrum of this galaxy is rich in 
high-ionization emission lines which are neither spatially nor spectrally
resolved.

\item[{\bf 6C1016+3637}] We aligned the slit through two 
plausible optical IDs marked on the radio map presented 
by Law-Green et al.\ (1995a). Our WHT spectroscopy detects continuum emission 
from both these objects but separate line emission from about half-way between 
them; both candidates IDs are therefore likely to be spurious, with the 
true ID having a Dec (1950.0) of $+36~37~46$. 
Since this ID is likely to lie along the radio axis it is possible that
we have missed some fraction ($\ltsimeq 50$ per cent) of the line emission. 
In Table~\ref{tab:summary} we have placed a limit on the
$K-$band magnitude of the true ID equal to the magnitude of the 
northern optical object, since this was detected at $\approx
3 \sigma$ in the near-IR image of Eales et al.\ (1997).

\item[{\bf 6C1017+3712}] We targeted the optical position 
given by Lilly (1989) confirming the redshift he
referenced to H. Spinrad ({\it priv. comm.}).

\item[{\bf 6C1031+3405}]
Our 1995 spectrum yields an unambiguous redshift for object `b' of
Eales \& Rawlings (1996) [the redshift given in their 
Table 1 is incorrect] and its strong emission lines, together with its 
location near the radio midpoint, mean that this is almost certainly the
correct radio galaxy ID. Object `a', an $R-$ band ID proposed by Eales (1985b),
was observed in 1994 and 
has a tentative redshift of 0.449 from a weak but definite line at $7255 ~ \rm \AA$, assumed to be [OIII]5007, and a possible
line at $\approx 5413 ~ \rm \AA$, assumed to be [OII]3727.

\item[{\bf 6C1042+3912}] The Ly$\alpha$ emission line from this object
is detected in the far blue, but two confirming emission
lines are also evident in the 2D blue-arm spectrum.

\item[{\bf 6C1045+3403}] The redshift of this object is based on five
emission lines.

\item[{\bf 6C1113+3458}] There is just one emission feature in the 
spectrum of this
object which because of its wavelength and huge equivalent width we take to be
Ly$\alpha$. The lack of any confirming features renders this redshift insecure.

\item[{\bf 6C1123+3401}]
There is only one definite emission line in the spectrum of this
object so its redshift is not yet secure.
In $F_{\lambda}$ the spectrum rises roughly 
$\propto \lambda^{4}$ from $\lambda \approx 5000 ~ \rm 
\AA$ until $\lambda \approx 7200 ~ \rm \AA$ then flattens off.
Together with the near-IR photometry (Eales et al.\ 1997), this produces
a spectral energy distribution that is 
reminiscent of the old red galaxies studied by Dunlop (1999);
by identifying the continuum flattening with a rest-frame wavelength
$\approx 3250 ~ \rm \AA$, we estimate a redshift $\sim 1.2$, which 
means that a plausible identification of the emission line
is CII]2326. At this redshift [OII]3727 would lie just beyond the wavelength
span of the 1995 WHT spectroscopy and
the relatively poor quality of the 1992 WHT (FOS-II) data means that the
lack of any detected feature is not a strong argument against the
veracity of the tentative redshift.

\item[{\bf 6C1129+3710}] 
Our WHT spectrum contains a
single bright emission line, but additional deeper 
optical spectroscopy 
(P. Best \& S. Rawlings, {\it priv. comm.}) detects additional lines 
which confirm the redshift obtained if this feature is associated with [OII]3727.

\item[{\bf 6C1134+3656}] 
Our WHT spectrum confirms the tentative
redshift obtained by Rawlings et al. (1990), although their claimed
CIII]1909 feature is not seen in the new spectrum.

\item[{\bf 6C1143+3703}] We observed this source in both 1994 and 1995
(with the same pointing position, but different slit PAs) but 
detected neither continuum nor lines: 
excepting the regions of the strongest night-sky emission lines,
we can put a $5 \sigma$ limit on any emission line at the 
$\sim 1 \times 10^{-19} ~ \rm W m^{-2}$ level.
Being only marginally detected at $K-$band, this object is the faintest 
object in the 6CE sample.

\item[{\bf 6C1158+3433}] The spectrum of this object is relatively poor because it
was a short exposure taken near twilight. There is a tentative detection of 
a single emission line which, if taken as [OII]3727, is consistent 
with a plausible $4000$-$\rm \AA$ spectral break.
The precise redshift of this object is therefore uncertain
but both the spectral shape and magnitude are those expected for a 
giant elliptical galaxy at $z \sim 0.5$. 

\item[{\bf 6C1159+3651}]
In 1994 we observed at a PA selected 
to include a companion object about 6 arcsec to the east (Lilly 1989) and in 
1995 at the PA of the radio source. In all the red-arm spectra we see
continuum from the ID amounting to $R \sim 24$, in 
reasonable agreement with the photometry
of Lilly, but there are no emission lines in the coadded data
to a $5 \sigma$ limit of $\sim 1 \times 10^{-19} ~ \rm W m^{-2}$
(excepting the regions of the strong night-sky emission lines).
The continuum is roughly flat in $F_{\lambda}$, although at the
signal-to-noise ratio of the detected continuum it is impossible
to search for any spectral breaks. Note the following: (i) 
the redshift given for this source in Table 1 of Eales \& Rawlings (1996) 
is almost certainly incorrect; and (ii) the near-IR photometry 
(Lilly 1989) may include some contribution from the companion object
(from which we detected featureless continuum in 1994).

\item[{\bf 6C1204+3708}] 
Since this is a large (51.5 arcsec) radio source, we began our blind exposure
with a 5-arcsec wide slit, and on seeing no emission lines in the first red
exposure, we widened the slit to 7 arcsec for 
the second red exposure, and more importantly 
(given the much higher ratio of read noise to sky noise in the blue)
for the remainder of the blue exposure. 
We detected strong emission lines in the blue-arm spectrum from a 
location along the slit consistent with the position of the
near-IR ID of Eales et al. (1997).
The line parameters for Ly$\alpha$ are rather uncertain
because a cosmic ray event effects two pixels near to its 
location. We also detected the object near the western hotspot in the
Eales et al.\ image: if the single emission line detected from this
object is [OII]3727, its redshift is $z=0.864$ making it 
foreground to the radio galaxy.

\item[{\bf 6C1204+3519}] This object provides a salutary lesson on the
dangers of assuming that a bright, highly-extended emission line in the blue
is always Ly$\alpha$: the detection of three clear emission lines leaves no
doubt that the brightest line, which is extended over 6 arcsec,
is CIV not Ly$\alpha$. The HeII and CIII] lines peak 
at the location of the continuum, and are also detected 
in a continuum-free blob 
$\approx 2$ arcsec (approximately) to the north;
the line fluxes of these two components are roughly equal, although there is
additional diffuse material, at least in CIV. In the 
red-arm spectrum there is also evidence for MgII emission
from the blob (although none is clearly detected in the total-light
spectrum of Fig.~\ref{fig:spectra}). There are hints in both the HeII and
CIII] emission that the blob is blue-shifted by 
$\approx 750 ~ \rm km ~ s^{-1}$ with respect to the 
continuum-emitting object, which might also help explain the broad
width of the CIV feature in the total-light spectrum.  
These observations might explain two unusual features of this object:
(i) relatively unusually for the 6CE sample, the near-IR emission
is well aligned with the radio axis (Eales et al. 1997), perhaps 
caused by two near-IR objects coinciding with the spatial peaks in the 
optical line emission; and (ii) the relatively large offset between the 
putative radio central component (Law-Green et al. 1995a)
and the position of the optical ID --- either the 
`radio core' is actually a knot in the jet, coincident with the
emission-line blob, or the active nucleus lies within the blob
approximately 2-arcsec to the north of an optically-brighter galaxy.

\item[{\bf 6C1212+3805}] The spectrum of this object shows only a single
strong emission line which we associate with 
[OII]3727 since the spectrum has red continuum detected on both sides of 
the line, and the spectral energy distribution 
rises rapidly into the near-IR 
(Eales et al. 1997; see also Maxfield et al. 1995) where the
ID is a bright as a typical 6CE galaxy at $z \sim 1$.
We eliminated the possibility that the single bright line is Ly$\alpha$
by detecting the ID in a deep ($3 \times 600 ~ \rm s$) $B-$band image with the WHT
Prime Focus Camera on 19 Apr.\ 1996 since this pass-band
would lie below the Lyman limit if the line were Ly$\alpha$.
The line emission is spatially resolved but not in an 
obviously asymmetric way with respect to the continuum. Our slit
also encompassed the companion galaxy to the north-west 
in the image of Eales et al.\ from which we detect 
spatially-extended red continuum but
no emission lines.

\item[{\bf 6C1217+3645}] 
Three highly tentative emission features are detected which line up well 
with a redshift supported by a far-from-convincing
$4000$-$\rm \AA$ break. The emission from this object is spatially unresolved in
seeing of about 1 arcsec, but although the signal-to-noise ratio is 
too low to be certain, there are no hints that the putative 
CIII] and MgII features are broad as would be the case if,
as suggested by Ben\'{i}tiz et al. (1995), this
object should be re-classified as a quasar. 
Note, further, that the 1.24-arcsec seeing, 
near-IR image of this galaxy is resolved
(Roche et al. 1998). 
The radio properties of this source are highly unusual (Best et al. 1999),
and one might speculate that the compact nature of the ID, and its 
relatively blue colour are further indications that this source
is anomalous within the sample, one possibility being that
it is a radio halo source in a rich cluster rather than a classical double.

\item[{\bf 6C1256+3648}]
Our WHT spectrum contains only a
single definite emission line, but additional deeper optical spectroscopy 
(P. Best \& S. Rawlings, {\it priv. comm.}) detects additional lines 
which confirm the redshift obtained if this feature is associated with [OII]3727.
The emission line region extends from the galaxy 
over the entire southern radio lobe;
no line emission is seen towards the northern lobe. 
The slit also encompassed a galaxy $\approx 17$ arcsec 
to the north-east of the radio galaxy, visible as a similar 
magnitude galaxy in the image of Eales et al. (1997): we detect an 
[OII] line in this galaxy at a similar redshift, so this galaxy is a 
companion to the radio galaxy in line with the 
suggestion of Lilly (1989) that this field is 
reminiscent of the core of a rich cluster.

\item[{\bf 6C1257+3633}]
The extension of the [OII]3727 emission is towards the south eastern 
radio lobe which shows very little radio polarization
(Best et al. 1999). The extended emission shows a distinct 
blob $\sim 3$ arcsec in this direction which, following
the near-IR/radio registration of Best et al., puts it 
slightly beyond a prominent knot in the radio emission. Although there
is no clear MgII emission in the total-light spectrum 
(Fig.~\ref{fig:spectra}), there
are hints of MgII emission in the 2D spectrum at the position of the
blob, but no hints of high-ionization lines
suggesting a lower effective ionization parameter for this emission.

\end{description}

\subsubsection{Notes on other 6CE sources}
\label{sec:obj_by_obj_lit}

\begin{description}

\item[{\bf 6C0825+3452}] Inspection of the 
NVSS image (Condon et al. 1998) of this radio source 
shows that there is a confusion problem which means that
the 151-MHz flux density $S_{151}$ of the main source is likely to
lie uncomfortably close to the 6CE sample flux density limit. 
In the absence of evidence to the contrary we 
assume that the main source still lies above 
the $S_{151}$ lower limit. There are 
two faint near-IR objects close to the position
of this radio source: Eales et al. (1997) favoured the brighter
south-eastern object as the ID, but the
subsequent detection of an inverted radio central component at the position
of the fainter north-western object (Best et al. 1999) means that this is
more likely to be the correct near-IR counterpart. The redshift
was obtained by blind spectroscopy.

\item[{\bf 6C1045+3553}]
The redshift is based on a single emission line presumed to be [OII].

\item[{\bf 6C1045+3513}]
The nature and redshift of the ID of this source reported by
Naundorf et al. (1992) is incorrect: spectroscopy shows that as 
originally suspected by Eales (1985b) the object is a quasar and lies at 
$z = 1.594$, not at $z \sim 0.7$. 
The cause of the confusion is the optical/near-IR colour of 
the object which is similar to that of an elliptical galaxy 
at $z \sim 0.7$: the compact near-IR image of Eales et al. (1997) 
shows that the red colour is the result of light reddening of a quasar nucleus
just like 3C318, the nature and redshift of which has recently 
similarly been revised (Willott, Rawlings \& Jarvis 2000a).

\item[{\bf 6C1255+3700}]
The redshift reported for this source by Warner et al. (1983) is incorrect -- 
we have estimated an observed equivalent width of 
$\approx 50 ~ \rm \AA$ for the [OIII]5007 line from the 
spectrum of Eracleous \& Halpern (1994) which also shows clear broad
H$\beta$.

\item[{\bf 6C1301+3812}]
No radio central component has been detected for this object, so,
as discussed by Allington-Smith, Lilly \& Longair (1985), identification
of the radio source with their galaxy `B' remains an unproven assumption.

\end{description}

\begin{figure*}
\begin{center}
\setlength{\unitlength}{1mm}
\begin{picture}(150,120)
\put(-20, -10){\includegraphics{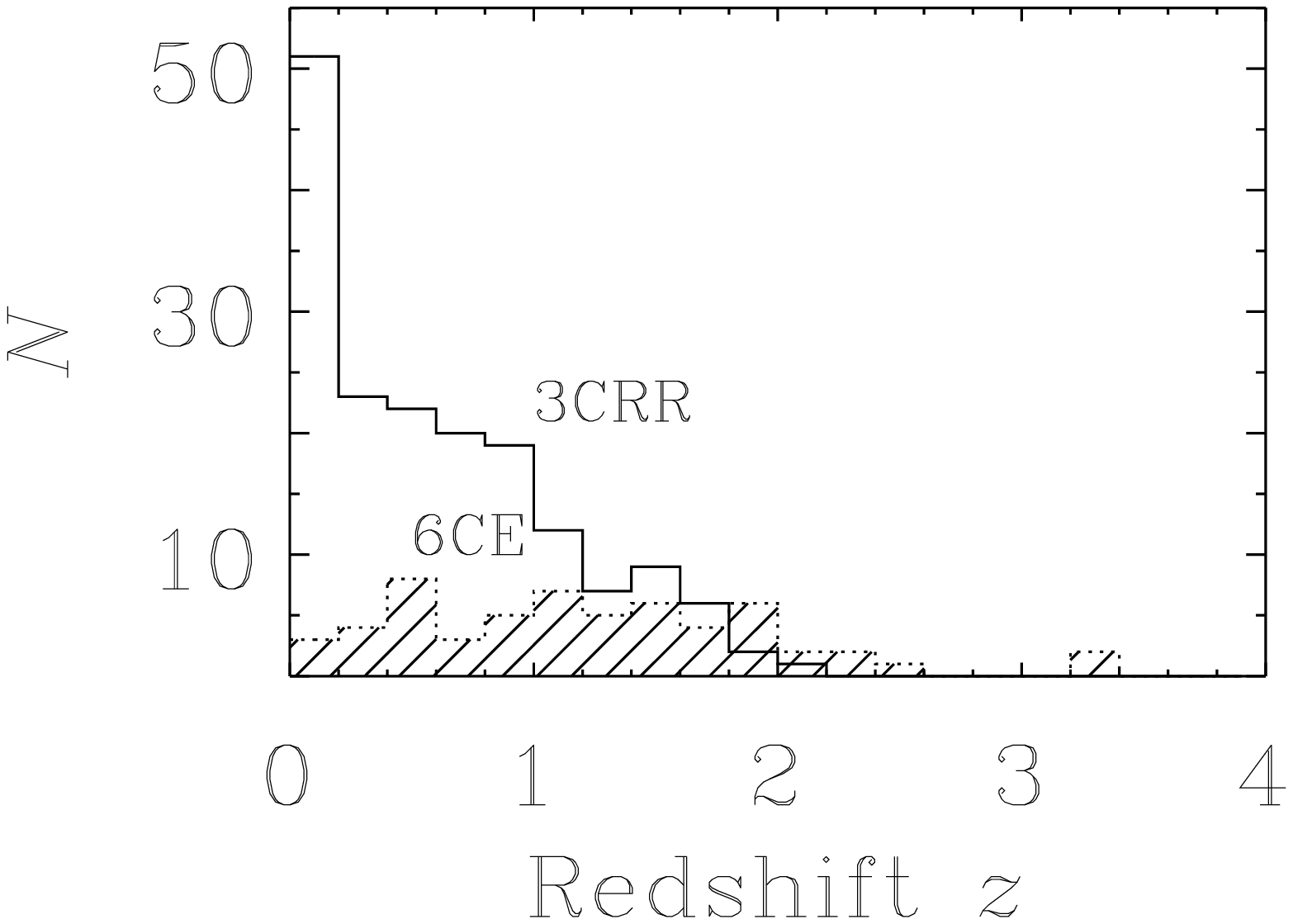}}
\end{picture}
\end{center}
{\caption[junk]{\label{fig:nz} Histograms of the redshift
distributions $N(z)$ of the 6CE sample (dotted line and
cross-hatched) and the
3CRR sample (full line and open). The bin width on 
each histogram is $\Delta z = 0.2$.
}}
\end{figure*}

\subsection{Objects without secure spectroscopic redshifts}
\label{sec:noz}

Inspection of Table~\ref{tab:summary} shows 
that, excluding the object obscured by a bright star,
only two 6CE objects 
(6C1143+3703 and 6C1159+3651)
lack any sort of spectroscopic redshifts.
This remarkable (97 per cent) level of redshift completeness 
is unprecedented at these flux density levels (e.g.\ McCarthy et al. 1996), 
and provides compensation for the relatively modest sample size. 
One of these troublesome objects (6C1159+3651) is relatively bright in the
near-IR, but faint in the optical, properties that are 
characteristic of a galaxy in the redshift range $0.8 \leq z \leq 2$
and we take $z=1.4$ here. The other object
(6C1143+3703) is faint at $K$ so, rather arbitrarily
given the scatter in the $K-z$ relation at faint
magnitudes (e.g.\ Eales et al. 1997), we take $z = 2$.

The 56 spectroscopic redshifts given 
in Table~\ref{tab:summary} are not yet all secure. In four cases 
(6C1045+3553, 6C1113+3458, 6C1123+3401, 6C1212+3805) this is because the
redshift depends on the correct identification of a single
emission line; in a further three
cases (6C0919+3806, 6C1158+3433, 6C1217+3645)
this is because there are no clear-cut spectral features in the
observed spectrum, but typically several tentative features 
consistent with a single redshift. Longer spectroscopic exposures 
of all these seven objects would be useful. In one further
case (6C1301+3812) there is some doubt about whether the
measured redshift is for the host galaxy of the 6C radio source.

\subsection{The redshift distribution}
\label{sec:nz}

The redshift distribution of the 6CE sample is compared with that of the 3CRR
sample in Fig.~\ref{fig:nz}. Note the very different redshift 
distributions resulting from the nature and redshift dependence of the 
radio luminosity function and the observable comoving volume
on our light cone (see Willott et al. 2000c). The median redshift of the
6CE sample is $\approx 1.1$ with a maximum redshift $\approx 3.4$, while
the 3CRR sample has a median redshift $\approx 0.55$
and a maximum redshift $\approx 2$.

\scriptsize
\begin{table*}

\begin{center}

\begin{tabular}{llllllr}

\hline
\hline

Object   & Date & \multicolumn{2}{c}{Position (1950.0)} &
Exposures & \multicolumn{2}{c}{Slit Width and PA} \\

& & & &
& (arcsec) & ($^{\circ}$) \\

\hline

6C0820+3642 & 27/01/95 & 08 20 33.89 & +36 42 28.3 (r) &
($1 \times 900$B,  $1 \times 900$R) & 3 & 155 (r) \\

\hline

6C0901+3551 & 08/01/94 & 09 01 25.10& +35 51 00.8 (r) &
($1 \times 1800$B, $2 \times 900$R) & 3 & 120 (r) \\

\hline

6C0905+3955 & 06/01/94 & 09 05 04.95 & +39 55 34.9 (c) &
($1 \times 1800$B, $2 \times 900$R) & 3 & 103 (i) \\
          & 08/01/94 &  09 05 04.95  & +39 55 34.9 (c) & 
($1 \times 1800$B, $2 \times 900$R) & 3 & 103 (i) \\

\hline

6C0919+3806 & 28/01/95 & 09 19 08.00 & +38 06 51.7 (c?) &
($2 \times 1800$B, $4 \times 900$R) & 2 & 151 (r)  \\

\hline

6C0930+3855 & 07/01/94 & 09 30 00.76  & +38 55 07.4 (r) &
($1 \times 1800$B, $2 \times 900$R) & 3 & 0 (r)  \\

\hline

6C0943+3958 & 27/01/95 & 09 43 13.67  & +39 58 11.2 (r) &
($1 \times 600$B, $1 \times 600$R) & 3 & 100 (r)  \\

\hline

6C1011+3632 & 07/01/94 & 10 11 16.91  & +36 32 12.4 (c) &
($1 \times 1500$B, $1 \times 900+ 1 \times 600$R) & 2 & 0 (r)  \\

\hline

6C1016+3637 & 07/01/94 & 10 16 58.45  & +36 37 49.7 (o) &
($1 \times 1800$B, $2 \times 900$R) & 3 & 167 (o)  \\

\hline

6C1017+3712 & 27/01/95 & 10 17 44.70  & +37 12 08.0 (o) &
($1 \times 600$B, $1 \times 600$R) & 3 & 45 (r)  \\

\hline

6C1031+3405 & 06/01/94 & 10 31 44.21 & +34 04 54.7 (o) & 
($1 \times 1800$B, $2 \times 900 $R) & 3 & 135 (r) \\
            & 27/01/95 & 10 31 44.29  & +34 04 57.9 (r) &
($1 \times 1800$B, $1 \times 1800$R) & 3 & 135 (r)  \\

\hline

6C1042+3912 & 07/01/94 & 10 42 23.73 & +39 12 25.2 (r) & 
($1 \times 1800$B, $2 \times 900 $R) & 3 & 110 (r) \\

\hline

6C1045+3403 & 06/01/94 & 10 45 24.18 & +34 03 37.9 (r) &
($1 \times 1800$B, $2 \times 900 $R) & 3 & 110 (r) \\

\hline

6C1113+3458 & 06/01/94 & 11 13 47.65 & +34 58 46.5 (r) &
($1 \times 1800$B, $2 \times 900 $R) & 3 & 18 (r) \\
            & 08/01/94 & 11 13 47.65 & +34 58 46.5 (r) &
($1 \times 1200$B, $1 \times 1200 $R) & 3 & 18 (r) \\

\hline

6C1123+3401 & 29/01/92 & 11 23 43.09 & +34 01 57.0 (r) &
($2 \times 900$B, $2 \times 900 $R) & 4 &  93  (v) \\
            & 28/01/95 & 11 23 43.09 & +34 01 57.0 (r) &
($1 \times 1800$B, $2 \times 900 $R) & 2 & 176 (v) \\
            & 31/01/95 & 11 23 43.09 & +34 01 57.0 (r) &
($1 \times 1800$B, $2 \times 900 $R) & 2 & 115 (v) \\

\hline

6C1125+3745 & 29/01/92 & 11 25 49.77 & +37 45 25.3 (o) &
($2 \times 900$B, $2 \times 900 $R) & 4 &  87  (r) \\

\hline

6C1129+3710 & 06/01/94 & 11 29 55.45 & +37 10 51.9 (r) &
($1 \times 1000$B, $1 \times 900 $R) & 3 & 138 (v) \\

\hline

6C1134+3656 & 08/01/94 & 11 34 28.60 & +36 56 30.0 (r) &
($1 \times 1800$B, $2 \times 900 $R) & 3 & 130 (r) \\

\hline

6C1143+3703 & 07/01/94 & 11 43 01.23& +37 03 07.4 (r) &
($1 \times 1800$B, $1 \times 1800 $R) & 2 &  68 (v) \\
            & 28/01/95 & 11 43 01.23& +37 03 07.4 (r) &
($1 \times 1550$B, $1 \times 1550 $R) & 2 &  90 (v) \\

\hline

6C1158+3433 & 08/01/94 & 11 58 19.50 & +34 33 28.9 (o) &
($1 \times 600$B, $1 \times 600 $R) & 3 & 90 (v) \\

\hline

6C1159+3651 & 09/01/94 & 11 59 20.94& +36 51 36.2 (r) & 
($1 \times 1800$B, $2 \times 900 $R) & 3 &  100 (o) \\
            & 27/01/95 & 11 59 20.94& +36 51 36.2 (r) & 
($1 \times 1800$B, $1 \times 1800 $R) & 2 &  32 (r) \\

\hline

6C1204+3708 & 09/01/94 & 12 04 22.08 & +37 08 17.7 (r) &
($1 \times 1200$B, $2 \times 600 $R) & 5-7 & 107 (r) \\

\hline

6C1204+3519 & 07/01/94 & 12 04 59.32 & +35 19 46.8 (o) &
($1 \times 1800$B, $2 \times 900 $R) & 2 & 12 (r) \\

\hline

6C1212+3805 & 27/01/95 & 12 12 26.04 & +38 05 31.0 (r) &
($1 \times 1800$B, $1 \times 1800 $R) & 2.7 & 130 (r) \\

\hline

6C1217+3635 & 09/01/94 & 12 17 40.16 & +36 45 45.4 (o) &
($1 \times 1200$B, $2 \times 600 $R) & 2 & 95 (o) \\
            & 27/01/95 & 12 17 40.16 & +36 45 45.4 (r) &
($1 \times 1800$B, $1 \times 1800 $R) & 1.8 & 40 (r) \\

\hline

6C1256+3648 & 06/01/94 & 12 56 44.75 & +36 48 08.4 (o) &
($1 \times 1800$B, $2 \times 900 $R) & 3 & 50 (r) \\

\hline

6C1257+3633 & 09/01/94 & 12 57 08.78 & +36 33 11.5 (c) &
($1 \times 1200$B, $2 \times 600 $R) & 3 & 138 (r) \\

\hline
\hline

\end{tabular}

\end{center}
\caption{\label{tab:journal}
Journal of optical spectroscopy of the 6CE targets.
The letter in brackets after the target position denotes its origin: 
`c' denotes a radio central component; `r', the
centre of an extended radio structure; `o', the position of the optical ID;
`i', the position of an infrared ID.
Exposures gives the time in s of each exposure in the blue (B) 
and red (R) arms. The letter in brackets after the PA 
denotes whether the slit alignment was chosen to match the radio (`r')
axis, or the major axis of infrared (`i') or optical (`o') structures, 
or whether the choice was made irrespective of the object properties (`v'), 
normally at an angle close to the parallactic angle at the time of observation.
}
\end{table*}

\begin{table*}

\begin{center}
  \begin{tabular}{llccrrrl}

\hline
\hline

Object     & Line & $\lambda_{rest}$ &  $\lambda_{meas}$ & 
Continuum & Flux (\% Error) & Extent & Comments \\ 
           &      & ($\rm \AA$)          &  ($\rm \AA$)          & 
(10$^{-22}$ W m$^{-2}$ $\rm \AA^{-1}$) & (10$^{-19}$ W m$^{-2}$) & (arcsec) & \\

\hline

6C0820+3642 & Ly$\alpha$ & 1216 & 3464 &$\leq10.0$ &15.0 (30) & 7  & FWHM unreliable \\
            & CIV        & 1549 & 4420 & 20.0   & 1.2 (50) &    & \\
            & HeII       & 1640 & 4691 & 25.0   & 3.3 (50) &    & near CR removed \\
            & CIII]      & 1909 & 5463 & 25.0   & 1.6 (50) &    & \\ 
\hline

6C0901+3551 & Ly$\alpha$ & 1216 & 3533 &$\leq10.0$&18.8 (15) & 1  & FWHM = 0-1300\\
            & CIV        & 1549 & 4498 & 10.0   & 2.6 (50) &    & \\
            & HeII       & 1640 & 4764 & 10.0   & 2.6 (30) &    & \\
            & CIII]      & 1909 & 5538 & 25.0   & 1.0 (50) &    & \\ 
\hline

6C0905+3955 & Ly$\alpha$ & 1216 & 3505 &$\leq10.0$& 8.4 (30) & 5  & FWHM = 0-1300\\
            & CIV        & 1549 & 4457 & 15.0   & 2.0 (40) &    & \\
            & HeII       & 1640 & 4719 & 20.0   & 0.5 (50) &    & \\

\hline

6C0919+3806 & CIII]?!    & 1909 & 5050 &  5.0   & 0.2 (50) &    & \\
            & CII]?!     & 2326 & 6174 &  5.0   & 0.5 (50) &    & \\

\hline

6C0930+3855 & Ly$\alpha$ & 1216 & 4116/4140 & $\leq5.0$ & 3.0 (30) &$<1$& Double-peaked \\

\hline

6C0943+3958 &[OII]       & 3727 & 7593 & 10.0   & 6.9 (15) & 4 & FWHM = 400-900 \\

\hline

6C1011+3632 &CIII]       & 1909 & 3899 & 20.0   & 1.3 (30) &      & \\
            &[NeIV]      & 2424 & 4955 & 20.0   & 1.0 (20) &      & \\
            &MgII        & 2798 & 5711 & 25.0   & 0.6 (50) &      & \\
            &[NeV]       & 3346 & 6835 & 20.0   & 0.5 (30) &      & \\
            &[NeV]       & 3426 & 6998 & 20.0   & 1.7 (15) &      & \\
            &[OII]       & 3727 & 7617 & 20.0   & 1.5 (20) & $<1$ & FWHM=0-400 \\
            &[NeIII]     & 3869 & 7904 & 20.0   & 2.0 (20) &      & \\

\hline

6C1016+3637 & Ly$\alpha$ & 1216 & 3516 & $\leq10.0$& 4.2 (20) & 4  & FWHM =0-1200\\
            & CIV        & 1549 & 4488 & $\leq10.0$& 1.0 (30) &    &            \\

\hline

6C1017+3712 &[NeV]       & 3426 & 7025 & 10.0   & 1.4 (50) &      & \\
            &[OII]       & 3727 & 7651 & 10.0   & 4.5 (30) &  5   & FWHM=750-1100 \\

\hline

6C1031+3405 & Ly$\alpha$ & 1216 & 3440 & 10.0   & 7.0 (30) & $<1$ & FWHM=0-800 \\
            & CIV        & 1549 & 4386 & 30.0   & 2.4 (30) &    & \\
            & HeII       & 1640 & 4646 & 25.0   & 1.0 (30) &    & \\
            & CIII]      & 1909 & 5404 & 20.0   & 1.7 (40) &    & \\ 
\hline

6C1042+3912 & Ly$\alpha$ & 1216 & 3368 &$\leq 10.0$& 8.5 (30) & $<1$ & FWHM=0-1100\\
            & CIV        & 1549 & 4291 &$\leq 10.0$& 1.0 (30) &      & \\
            & HeII       & 1640 & 4544 &$\leq 10.0$& 0.6 (50) &      & \\

\hline

6C1045+3403 & Ly$\alpha$ & 1216 & 3435 &$\leq 10.0$&25.8 (15) &  5   & FWHM=800-1800\\
            & CIV        & 1549 & 4382 & 15.0      & 1.5 (30) &      & \\
            & HeII       & 1640 & 4635 & 15.0      & 1.2 (50) &      & \\
            & CII]       & 2326 & 6567 & 10.0      & 0.6 (60) &      & \\
            &[NeIV]      & 2424 & 6835 & 20.0      & 0.6 (50) &      & \\

\hline
\hline

\end{tabular}

\end{center}
  \caption{\label{tab:measurements}
Measurements obtained from the WHT spectra
of 6CE radio sources. The `?' denotes an uncertain line
identification and the `!' denotes a feature which is plausibly spurious.
Errors on the line fluxes represent $\approx$90\%
confidence intervals expressed as a percentage of the
best line flux estimate; for the strongest lines these are dominated
by roughly equal contributions from uncertainties in fixing 
the local continuum level, and from the absolute flux calibration
(and plausible slit losses). 
The spatial extents of the emission lines were estimated by
evaluating the full-width zero-intensity of a cross-cut
through the emission line, deconvolved from the seeing.
Line widths (given in units of
km s$^{-1}$) were estimated from the FWHM of the best
Gaussian fit to each line, the lower value of a range 
assumes the line-emitting region fills the slit, and the 
higher value assumes that it is broadened only by the seeing.
`CR contam' is short-hand for cosmic-ray contamination.
}
\end{table*}
\normalsize

\clearpage

\addtocounter{table}{-1}

\scriptsize

\begin{table*}

\begin{center}
  \begin{tabular}{llccrrrl}

\hline
\hline

Object     & Line & $\lambda_{rest}$ &  $\lambda_{meas}$ & 
Continuum & Flux (\% Error) & Extent & Comments \\ 
           &      & ($\rm \AA$)          &  ($\rm \AA$)          & 
(10$^{-22}$ W m$^{-2}$ $\rm \AA^{-1}$) & (10$^{-19}$ W m$^{-2}$) & (arcsec) & \\

\hline

6C1113+3458 & Ly$\alpha$?& 1216 & 4142 & $\leq10.0$& 2.2 (20) & 3  & FWHM = 0-900\\

\hline

6C1123+3401 & CII]?      & 2326 & 5227 &  6.0 &  0.4 (50) &   &           \\

\hline

6C1125+3745 & CIV        & 1549 & 3471 & 77.0 & 48.0 (30) & $<1$ & FWHM $\sim$ 10000\\
            & CIII]      & 1909 & 4237 & 80.0 &  9.0 (50) &   &           \\
            & MgII       & 2798 & 6249 & 65.0 & 15.0 (25) &   &           \\

\hline

6C1129+3710 & [OII]      & 3727 & 7678 & 45.0 &  8.2 (30) & 7 & FWHM = 0-700\\
            & [NeIII]?!  & 3869 & 7974 & 45.0 &  1.2 (60) &   &           \\

\hline

6C1134+3656 & Ly$\alpha$ & 1216 & 3800 &$\leq 20.0$&13.0 (30) &  15   & FWHM=1000-1700\\
            & CIV        & 1549 & 4839 & 25.0      & 2.0 (40) &       & \\
            & HeII       & 1640 & 5121 & 30.0      & 2.6 (40) &       & \\

\hline

6C1158+3433 & [OII]?!    & 3727 & 5704 & 65.0 &  2.5 (50) & $<1$ & FWHM unreliable\\

\hline

6C1204+3708 & Ly$\alpha$ & 1216 & 3379 & 20.0      &47.0 (40) &  4   & CR contam\\
            & CIV        & 1549 & 4305 & 35.0      &11.5 (25) &      & FWHM=1200-1400\\
            & HeII       & 1640 & 4635 & 20.0      & 3.7 (30) &      & \\
            & CIII]      & 1909 & 5308 & 25.0      & 2.1 (40) &      & \\

\hline 

6C1204+3519 & CIV        & 1549 & 3681 & 20.0   &  4.5 (25) &  6 & FHWM=2000-2500  \\
            & HeII       & 1640 & 3894 & 25.0   &  1.5 (50) &    & CR contam? \\
            & CIII]      & 1909 & 5404 & 20.0   &  1.6 (40) &    &            \\ 

\hline

6C1212+3805 & [OII]?     & 3727 & 7256 & 10.0   &  3.0 (20) &  2 & FWHM=550-950\\

\hline

6C1217+3645 & CIII]?!    & 1909 & 3965 & 30.0   &  0.7 (60) &$<1$& CR contam  \\
            & MgII?!     & 2798 & 5830 & 25.0   &  0.7 (60) &    &            \\
            & [OII]?!    & 3727 & 7793 & 30.0   &  0.9 (60) &    &            \\
\hline

6C1256+3648 & [OII]      & 3727 & 7926 & 20.0   &  3.0 (40) &  6 & FWHM=0-500\\
            & [NeIII]?!  & 3869 & 8232 & 25.0   &  1.0 (60) &    &           \\

\hline

6C1257+3633 & CIII]      & 1909 & 3825 & 20.0   &  2.9 (40) &    &           \\
            & [NeIV]     & 2424 & 4855 & 15.0   &  1.3 (50) &    &           \\
            & [NeV]      & 3346 & 6705 & 40.0   &  0.5 (60) &    &           \\
            & [NeV]      & 3426 & 6862 & 35.0   &  1.2 (50) &    &           \\
            & [OII]      & 3727 & 7465 & 35.0   &  6.4 (20) &  5 & FWHM=0-500\\
            & [NeIII]    & 3869 & 7748 & 40.0   &  3.2 (30) &    &           \\

\hline
\hline

\end{tabular}

\end{center}
  \caption{{\bf (cont).}}
\end{table*}
\normalsize

\clearpage

\subsection{Emission line properties}
\label{sec:disco}

At redshifts $> 1.75$ the only 3CRR objects known are,
in terms of their radio luminosity, amongst the most extreme objects on our
light cone: the 16 $z > 1.75$ 6CE objects are typically 
$\sim 1$ dex less radio luminous.
Thirteen of these 16 objects lie in very restricted bands
of luminosity and and redshift, $1.75 \leq z \leq 2.45$ and 
$27.3 \leq \log_{10} (L_{\rm 151}) < 27.8$, and we will take
their optical properties to be characteristic of 
`high-redshift 6CE sources'. 
The two extreme luminosity (and redshift) 6CE sources 
have been studied in detail elsewhere (6C0902+3419 by, for example,
Lilly 1988, Eales et al. 1993b and
Carilli 1995; and 6C1232+3942 by Eales et al. 1993a), 
and the lowest luminosity $z > 1.75$ object is distinct in
its radio properties (its projected linear size is more than 1 dex lower
than all other 6CE objects st $z > 1.75$) and
as discussed in Sec.~\ref{sec:noz}, is one of the two 6CE sources
without a spectroscopic redshift.

\begin{figure*}
\begin{center}
\setlength{\unitlength}{1mm}
\begin{picture}(150,100)
\put(160,0){\includegraphics{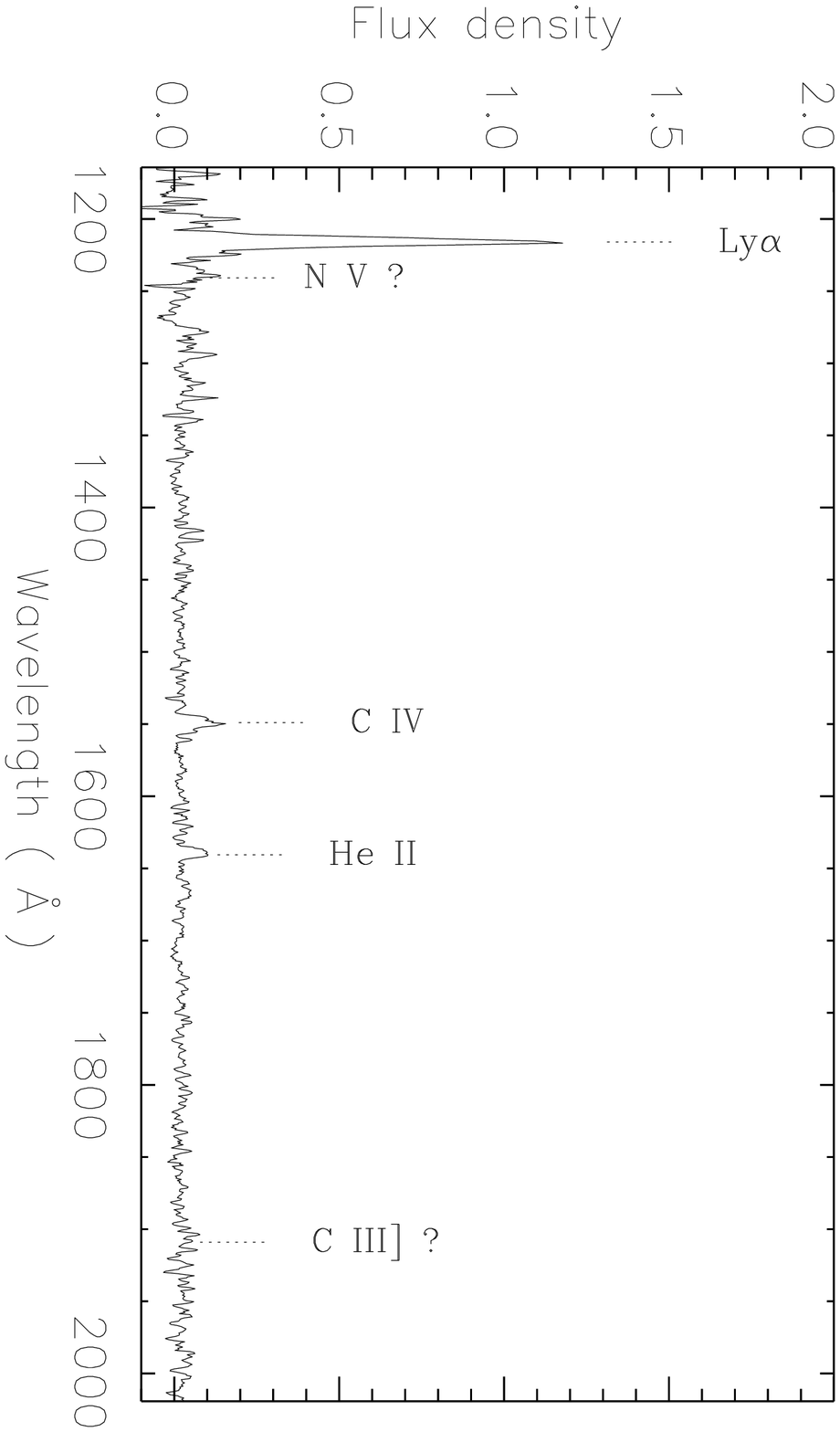}}
\end{picture}
\end{center}
{\caption[junk]{\label{fig:compo} 
Composite spectrum resulting from the combination of the 
11 spectra of $z > 1.75$ 6CE galaxies
discussed in Sec.~\ref{sec:spectra} with $z > 1.75$;
the method of combination is described in Sec.~\ref{sec:disco},
and the resultant spectrum has been smoothed with a 3-bin
boxcar filter. The flux density is measured in units
of $\rm W ~ m^{-2} ~ \AA^{-1}$ with an arbitrary normalization. 
}}
\end{figure*}

As well as spanning narrow ranges of $L_{151}$ and $z$,
the thirteen high-redshift 6CE sources also have strikingly similar
optical properties. Excepting 6C0824+3535, the objects
are narrow-line radio galaxies rather than quasars, 
and for the radio galaxies the
distribution in Ly$\alpha$ luminosities $L_{\rm Ly \alpha}$
is very narrow: from Table~\ref{tab:summary} we find
$\log_{10} L_{\rm Ly \alpha} = 36.45 \pm 0.36$. The tight spread
in $\log_{10} L_{\rm Ly \alpha}$ is presumably a manifestation of the
well-known correlation between narrow-line and radio luminosities
(e.g.\ Willott et al. 1999). Eleven of the twelve relevant 6CE radio galaxy
spectra are presented in this paper (Sec.~\ref{sec:spectra}) and their
apparent homogeneity encouraged us to create a composite
spectrum which is hopefully representative of 
a high-$z$ 6CE radio source. This composite was
constructed by: (i) changing the wavelength scale of each spectrum to
place each object at redshift zero, ensuring each had the same
binning in wavelength; (ii) scaling each de-redshifted 
spectrum to equalize the Ly$\alpha$ fluxes; and (iii) taking a bin-by-bin
average with an iterative sigma-clipping rejection algorithm
(with the clip set at $\pm 1.5 \sigma$).
The composite is shown in Fig.~\ref{fig:compo}; the line
fluxes relative to Ly$\alpha$ of CIV and HeII are 
0.13 and 0.08 respectively, and there are marginally significant
detections of the NV and CIII] lines. 
The CIII] line is clearly detected in at least
four of the objects making up the composite, and its low signal-to-noise
ratio in Fig.~\ref{fig:compo} is probably partially due to its 
location (at high redshift) within the spectral region of the 
dichroic cross-over for the WHT
observations. The rest-frame 
equivalent widths of Ly$\alpha$, CIV and HeII are within a factor of
two of $450$, $70$ and $30 ~ \rm \AA$ respectively.
These results are in good
quantitative agreement with the composite spectrum presented
by McCarthy (1993) which, since it was calculated from 
an independent but similar dataset (3C galaxies plus 
radio galaxies with similar low frequency flux densities to the
6CE sample), shows that there does appear to
be such a thing as a generic UV spectrum of radio-luminous high-redshift 
galaxies.

\subsection{Quasar fraction}
\label{sec:qf}

The low fraction of quasars in the 6CE sample was 
first noticed and discussed by Eales (1985b,c).
There are only three quasars out of the 18 6CE objects
in the $27.3 \leq \log_{10} (L_{\rm 151}) < 27.8$ luminosity bin
whereas amongst the 41 similarly radio-luminous 3CRR sources,
between 14 and 16 
(depending on how one classifies 3C22 and 3C41; see
Simpson, Rawlings \& Lacy 1999)
are quasars, i.e.\ a fraction $\approx 0.4$. 
Using a combined 3CRR/6CE/7CRS dataset, Willott et al. (2000b)
have argued that this fraction is virtually
independent of $L_{\rm 151}$ and $z$ above a critical 
value of $\log_{10} (L_{\rm 151}) \approx 26.5$. 
Taking 0.4 as the probability that a given source is viewed as a quasar, 
the Binomial probability of finding three or fewer 6CE quasars out of 
18 6CE objects is about 3 per cent. 

One possible reason 
for this (marginally significant) deficit of quasars
is that at the higher redshifts probed by the 
6CE sample (see Fig.~\ref{fig:pz})
quasar nuclei may be more easily hidden by dust:
perhaps partly because optical spectra probe shorter rest-frame
wavelengths, and perhaps partly because of an intrinsically
larger fraction of reddened lines-of-sight at high redshift 
(e.g.\ Willott et al. 2000a). Just one of the luminous 6CE
radio galaxies needs to be a reddened quasar for the binomial
probability of the low quasar fraction in 6CE to 
rise to $\sim 10$ per cent. Note also that, at the highest radio luminosities
and redshifts, one of the two 6CE galaxies is 6C0902+3419 which,
as pointed out by Carilli (1995), has radio properties which
are consistent with a jet orientation as close to the line-of-sight
as a lower-redshift 3CRR quasar. We conclude that the low quasar fraction
observed in the 6CE sample, although still plausibly an artefact of
small number statistics, hints at one or more
redshift-dependent effects that warrant further
investigation in a larger sample of high-$z$ radio sources.

\section*{Acknowledgements}

We are extremely grateful to a large number of collaborators who have,
over the years, devoted effort and observing time to the quest for
complete redshift information for the 6CE sample, namely: 
Jeremy Allington-Smith, Philip Best, Mark Dickinson, 
Chris Haniff, Steve Maddox, Richard Saunders, Steve Serjeant, Hy Spinrad
and Steve Warren. We thank Katherine Blundell, 
Duncan Law-Green, Patrick Leahy, Tony Lynas-Gray and Chris Willott for their 
practical help compiling data on the 6CE sample.
We also thank an anonymous referee for comments which helped to improve the
organization of this paper.
The William Herschel Telescope (WHT) is operated
on the island of La Palma by the Isaac Newton Group in the Spanish
Observatorio del Roque de los Muchachos of the Instituto de
Astrofisica de Canarias.
This research has made use of the NASA/IPAC Extragalactic Database, which
is operated by the Jet Propulsion Laboratory, Caltech, under contract
with the National Aeronautics and Space Administration.

\end{document}